\documentclass[11pt,preprint]{aastex}

\usepackage{color}
\usepackage{natbib}
\usepackage{units}
\usepackage{amsmath}
\tightenlines
\slugcomment{}

\newcommand \gas        {{\rm gas}}
\newcommand \Evib       {E_{\rm vib}}

\newcommand \Jpara     {J_{\parallel\bB}}
\newcommand \Jperp     {J_{\perp\bB}}

\newcommand \Tvib       {T_{\rm vib}}

\newcommand \bomega     {{\boldsymbol{\omega}}}

\newcommand \crit       {{\rm crit}}
\newcommand \abs        {{\rm abs}}

\newcommand \Angstrom   {\,{\rm \AA}}

\newcommand \bahat      {\hat{\bf a}}
\newcommand \bB         {{\bf B}}

\newcommand \bE         {{\bf E}}
\newcommand \behat      {\hat{\bf e}}

\newcommand \bJ         {{\bf J}}

\newcommand \bM         {{\bf M}}

\newcommand \beq        {\begin{equation}}
\newcommand \beqa	{\begin{eqnarray}}

\newcommand \cm         {\,{\rm cm}}

\newcommand \eeq	{\end{equation}}
\newcommand \eeqa	{\end{eqnarray}}

\newcommand \erg	{\,{\rm erg}}
\newcommand \Erot       {E_{\rm rot}}

\newcommand \gm         {\,{\rm g}}
\newcommand \GHz        {\,{\rm GHz}}
\newcommand \gtsim	{\gtrsim}		 
\newcommand \Ha 	{{\rm H}}

\newcommand \K  	{\,{\rm K}}

\newcommand \ltsim	{\lesssim}		 

\newcommand \mH         {m_{\rm H}}

\newcommand \muG        {\mu{\rm G}}

\newcommand \nH         {n_{\rm H}}

\newcommand \s	        {\,{\rm s}}

\newcommand \xtimes     {{\!\,\times\!\,}}

\newcommand{\oldtext}[1]{}

\countdef\decade=200
\advance\decade by \year
\countdef\hours=201
\advance\hours by \time
\divide\hours by 60
\countdef\mins=202
\advance\mins by \hours
\multiply\mins by 60
\multiply\hours by 100
\countdef\miltime=203
\advance\miltime by \hours
\advance\miltime by \time
\advance\miltime by -\mins

\begin{document}

\title{
        \vspace*{-3.0em}
        {\normalsize\rm submitted to {\it The Astrophysical Journal} 
        (2016 May\ 20)}\\ 
        \vspace*{1.0em}
        Quantum Suppression of Alignment in Ultrasmall Grains:\\
        Microwave Emission from Spinning Dust will be Negligibly Polarized
	}

\author{
B.\ T.\ Draine
\\
Princeton University Observatory, Peyton Hall, Princeton, NJ 08544-1001, USA;
draine@astro.princeton.edu}
\author{Brandon S.\ Hensley
\\
Jet Propulsion Laboratory, California Institute of Technology, 4800 Oak Grove
Drive, Pasadena CA 91109, USA
}
\begin{abstract}
The quantization of energy levels in very nanoparticles
suppresses
dissipative processes that convert grain rotational
kinetic energy into heat.
For grains small enough to have
$\sim$GHz rotation rates, the suppression of dissipation can be extreme.
As a result, alignment of such grains is suppressed.
This applies both to alignment of the grain body with its angular
momentum $\bJ$, and to alignment of $\bJ$ with the local magnetic
field $\bB_0$.
If the anomalous microwave emission is rotational emission from
spinning grains, it will be negligibly polarized at GHz frequencies,
with $P\ltsim 10^{-6}$ at $\nu > 10\GHz$.
\end{abstract}
\keywords{dust}

\section{Introduction
         \label{sec:intro}}

The emission from the interstellar medium (ISM) in the
Milky Way and other star-forming galaxies
includes strong mid-IR emission features at 3.3, 6.2, 7.7, 8.6, 11.3, 12.6,
and 17$\micron$ \citep[see, e.g.,][]{Smith+Draine+Dale+etal_2007}.
The only viable explanation for this emission is
a substantial interstellar population
of nanoparticles with the composition of polycyclic
aromatic hydrocarbons (PAHs), containing as few as $\sim$40 atoms
\citep{Tielens_2008}.
The PAHs have been identified by their characteristic IR emission
features, but it is possible that nanoparticles with other compositions --
such as silicates or metallic Fe -- could also be abundant.
For the densities and temperatures present in the ISM, 
nanoparticles containing fewer than $\sim$$10^3$ atoms
will inevitably be spinning at $\sim$GHz frequencies.

The so-called anomalous microwave emission (AME) observed
at 10--60$\GHz$ was interpreted as rotational emission from
rapidly-rotating nanoparticles
\citep{Draine+Lazarian_1998a,Draine+Lazarian_1998b}.
Given the PAH abundances and size distribution
required to explain the observed mid-IR emission,
it was natural to consider spinning PAHs as the source for the AME.
However, a recent observational study 
\citep{Hensley+Draine+Meisner_2016} failed to
find the expected correlation of AME emission with PAH abundance.
\citet{Hensley+Draine+Meisner_2016} therefore suggested that
spinning non-PAH (e.g., silicate or iron)
nanoparticles may also be present in the ISM.
Possible emission from silicate and iron nanoparticles has been
further discussed by
\citet{Hoang+Vinh+Lan_2016}, \citet{Hoang+Lazarian_2016a},
and \citet{Hensley+Draine_2016a}.

This paper examines the dynamics of dissipation in spinning interstellar
nanoparticles, whether composed of hydrocarbons, silicates, or other materials.
Two types of dissipation are discussed.  One is the internal dissipation
that allows a tumbling grain to minimize its rotational
kinetic energy by aligning $\bahat_1 =$ the principal axis
of largest moment of inertia with its angular momentum $\bJ$.
The other is the dissipation that occurs in a static magnetic field $\bB_0$
when $\bJ$ is not aligned with $\bB_0$.
In both cases, rotational kinetic energy is converted to heat.

Dissipative processes in grains 
have usually been treated in the classical limit where
the rotating body has many internal degrees of freedom.
However, in very small grains, energy level quantization
will
suppress intramolecular vibration-rotation energy transfer
(IVRET)
and dissipation of rotational energy.

Here we examine the quantum suppression
of dissipation in spinning grains.
Vibration-rotation energy exchange must be suppressed
when the vibrational energy level spacing $\Delta E$
is larger than the intrinsic width $\delta E$ of the energy levels.
We estimate the suppression factor,
as a function of the grain's size and vibrational energy
content $\Evib$.

We calculate the implications of this quantum suppression on both
the alignment of $\bJ$ with $\bB_0$, and on the alignment 
of the principal axis $\bahat_1$ with $\bJ$.
Quantum suppression effects
are extreme for the smallest nanoparticles, leading to
almost total suppression of alignment of $\bJ$ with $\bB_0$ for the
smallest (and therefore most rapidly-rotating) grains.
If the AME is rotational emission from nanoparticles, it
will be essentially unpolarized.

The paper is organized as follows.
Section \ref{sec:energy levels} reviews the energy levels of spinning
nanoparticles, and section \ref{sec:g_E} examines the
distribution of vibrational modes and energy levels.
Section \ref{sec:quantum suppression} estimates the factor $\psi_{\rm q}(\Evib)$
by which IVRET will be suppressed in
a nanoparticle, as a function of the vibrational energy $\Evib$
present in the nanoparticle.
Section \ref{sec:align a with J}
concerns the quantum suppression of alignment of the grain body with $\bJ$.
In section \ref{sec:Jchanging} we discuss the quantum suppression of
magnetic
dissipation in either paramagnetic or ferromagnetic grains.
The rotation and alignment of spinning nanoparticles is calculated in
section \ref{sec:excitation and alignment}.
In section \ref{sec:rotpol} 
we calculate the polarization of rotational emission from spinning
nanoparticles, as a function of frequency.  For conditions
characteristic of neutral diffuse clouds, we show that the
rotational emission at frequencies $>1\GHz$ should have very small
polarization, $\ltsim 0.01\%$.
In section \ref{sec:pol opt-IR} we calculate the degree of polarization
of thermal emission from spinning nanoparticles, and the
dichroic extinction contributed by such particles.
The results are discussed in section \ref{sec:discussion}, and
summarized in section \ref{sec:summary}.

\section{\label{sec:energy levels}
         Energy Levels of a Spinning Nanoparticle}
\subsection{Rotation}
Consider grains that can be approximated by spheroids,
with $I_\parallel,I_\perp,I_\perp$ being the eigenvalues of the moment
of inertia tensor.
Consider the case $I_\parallel>I_\perp$ (i.e., oblate spheroids).
Let $J=$ the total angular momentum quantum number.
In the center-of-mass frame, the total energy of the grain is
\beqa
E_{v,J,K}&=&E_{v,0,0}+ E_{\rm rot}
\\ 
E_{\rm rot} &=& hc\left[B_v J (J+1) - (B_v-A_v)K^2\right]
~~~,
\eeqa
where the ``rotation constants'' $A_v$ and $B_v$ may depend on the
vibration state $v$, and the quantum number $K$ is the projection of
$\bJ$ along the symmetry axis $\bahat_1$.
For a spheroid, the rotation constants are
\beqa
A &=& \frac{\hbar}{4\pi c I_\parallel}
\\
B &=& \frac{\hbar}{4\pi c I_\perp}
~~~,
\eeqa
where $I_\parallel,I_\perp$ are the moments of inertia
for rotation parallel or perpendicular to the symmetry axis.
An oblate spheroid ($I_\parallel>I_\perp$) has $B>A$.

\begin{figure}[ht]
\begin{center}
\includegraphics[angle=0,width=8.0cm,
                 clip=true,trim=0.5cm 5.0cm 0.5cm 2.5cm]
{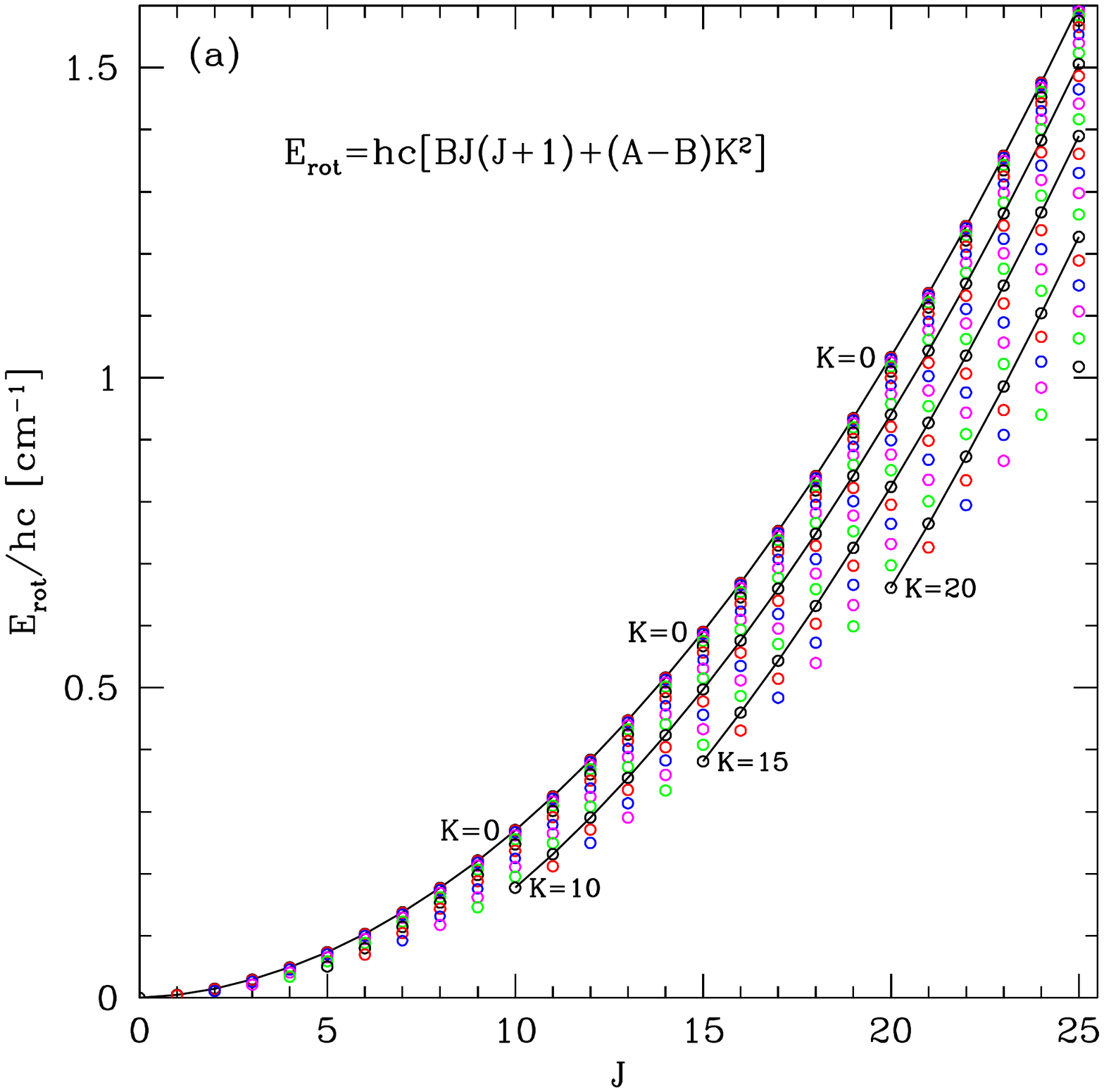}
\includegraphics[angle=0,width=8.0cm,
                 clip=true,trim=0.5cm 5.0cm 0.5cm 2.5cm]
{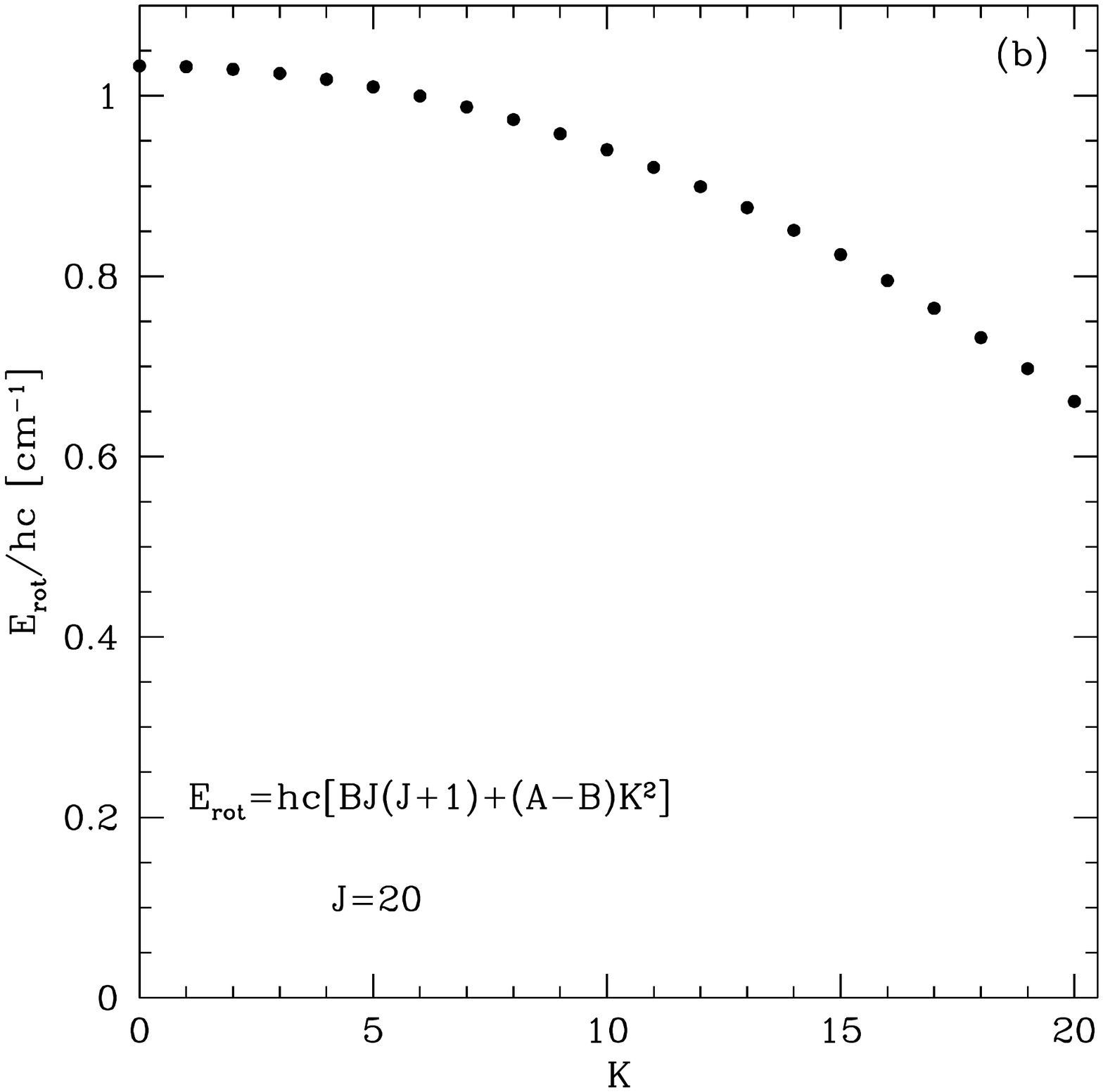}
\caption{\label{fig:Erot}\footnotesize
         Rotational kinetic energy for an oblate spheroid
         with axial ratio $b/a=2$, mass 
         800\,amu, and density $3.2\gm\cm^{-3}$.
         (a) $\Erot$ vs.\ $J$, for different values of $K$.
         (b) $\Erot$ vs.\ $K$, for $J=20$.}
\end{center}
\end{figure}
Figure \ref{fig:Erot}a shows the rotational states 
of an oblate spheroid with $J\leq25$.
For a given $J$, there are $J+1$ possible values of $K$,
with $K=0$ giving the highest energy, and $K=J$ giving
the lowest energy.
Figure \ref{fig:Erot}b shows the energy levels for $J=20$.

\subsection{Vibration}

Suppose the grain has total internal energy $E_{v,J,K}$ in vibration
and rotation.
The uncertainty 
$\delta E$ of the internal energy
is determined by the level lifetime.
If $A_{\rm rad}$ is the probability per unit time of a spontaneous
radiative transition, $\dot{N}_{\rm abs}$ is the probability per unit
time of absorbing a photon, 
and $\dot{N}_{\rm coll}$ is the probability
per unit time of an inelastic collision with a gas particle,
then
\beq
\delta E \approx \hbar(A_{\rm rad} + \dot{N}_{\rm abs} + \dot{N}_{\rm coll})
~~~.
\eeq
Radiation can be either purely rotational, or rovibrational; 
we write $A_{\rm rad}=A_{J\rightarrow J-1} + A_{\rm vib}$.
The Einstein A coefficient for pure rotational transitions is
\beq
A_{J\rightarrow J-1} \approx \frac{2\omega^3}{3\hbar c^3}\mu_\perp^2
~~~,
\eeq
where $\omega\approx 4\pi c B_v J$, and $\mu_\perp$ is the rms
electric dipole moment perpendicular to $\bJ$ for the spinning grain.
If $\nu_{\rm rot} = \omega/2\pi$, then
\beq
A_{J\rightarrow J-1} \approx 4\times10^{-6}
\left(\frac{\mu_\perp}{5\,{\rm D}}\right)^2
\left(\frac{\nu_{\rm rot}}{30\GHz}\right)^3\s^{-1}
~~~.
\eeq
Consider for the moment a spherical particle of
radius $a$.
At long wavelengths, the absorption cross section for
interstellar amorphous silicate grains is
\citep{Draine+Hensley_2016b}
\beq
C_\abs(\lambda) \approx 9\times10^{-19}\cm^2 
\left(\frac{100\micron}{\lambda}\right)^2 \left(\frac{a}{10^{-7}\cm}\right)^3
~~~;
\eeq
for this cross section, a grain with vibrational temperature
$\Tvib$ radiates photons at a rate
\beq
A_{\rm vib} 
= 4\pi \int C_\abs(\nu) \frac{B_\nu(T)}{h\nu} d\nu
\approx 2.6 \left(\frac{\Tvib}{10^2\K}\right)^5 
\left(\frac{a}{10^{-7}\cm}\right)^3 \s^{-1}
\eeq
and power
\beq
\dot{E}_{\rm vib} = 4\pi \int C_\abs(\nu) B_\nu(T) d\nu \approx 
1.9\xtimes 10^{-13} \left(\frac{\Tvib}{10^2\K}\right)^6 
\left(\frac{a}{10^{-7}\cm}\right)^3 
\erg\s^{-1}
~~~.
\eeq
In the interstellar radiation field,
the nanoparticle absorbs starlight photons at a rate
\citep[see Figure 11 of][]{Draine+Li_2001}
\beq \label{eq:Ndotabs}
\dot{N}_{\rm abs} \approx 5\xtimes10^{-7}U 
\left(\frac{a}{10^{-7}\cm}\right)^3 \s^{-1}
~~~,
\eeq
where the dimensionless factor $U$ is the 
intensity of the radiation field relative to the local interstellar
radiation field \citep{Mathis+Mezger+Panagia_1983}.

In gas of density $\nH$ and temperature $T_\gas$, the gas-grain
collision rate is
\beq
\dot{N}_{\rm coll} \approx \nH 
\left(\frac{8kT_\gas}{\pi \mH}\right)^{1/2}\pi a^2
= 1.4\xtimes10^{-7}\s^{-1}
\left(\frac{\nH}{30\cm^{-3}}\right)
\left(\frac{T_\gas}{10^2\K}\right)^{1/2} 
\left(\frac{a}{10^{-7}\cm}\right)^2
~~~.
\eeq
Thus for a silicate nanoparticle with $a\approx 5\times10^{-8}\cm$, 
we have a level lifetime
\beq
\tau \approx 
\left[
0.3 \left(\frac{\Tvib}{10^2\K}\right)^5 + 
4\xtimes10^{-6} \left(\frac{\nu_{\rm rot}}{30\GHz}\right)^3
+ 6\xtimes10^{-8}U
+ 3\xtimes10^{-8}\left(\frac{\nH}{30\cm^{-3}}\right)
\left(\frac{T_\gas}{10^2\K}\right)^{1/2}
\right]^{-1}\s
\eeq
corresponding to a level width
\beq
\frac{\delta E}{hc} \approx 1.6\xtimes10^{-12}
\left[\left(\frac{\Tvib}{10^2\K}\right)^5 + 
10^{-5}\left(\frac{\nu_{\rm rot}}{30\GHz}\right)^3 +
2\xtimes10^{-7}U+
10^{-7}\left(\frac{\nH}{30\cm^{-3}}\right)
\left(\frac{T_\gas}{10^2\K}\right)^{1/2}\right]\cm^{-1}
~~~.
\eeq
The level lifetimes and widths are strongly dependent on the
vibrational temperature $\Tvib$.  Broadening
due to collisions, microwave rotational emission, and
starlight absorption are of
secondary importance so long as the grain has vibrational energy
content corresponding to $\Tvib\gtsim 20\K$.

\section{\label{sec:g_E}
         Vibrational Density of States}

A grain with $N$ atoms, in its electronic ground state, 
has $3N-6$ vibrational degrees of
freedom, each with vibrational quantum number $\tilde{v}_j$.
The vibrational state of the grain is specified by the list of
vibrational quantum numbers $v=\{\tilde{v}_1,...,\tilde{v}_{3N-6}\}$
of the modes.
Let $N_v(E)$ be the number of distinct vibrational states
$v=\{\tilde{v}_1,...,\tilde{v}_{3N-6}\}$ with total
vibrational energy $\Evib<E$.
If the vibrational modes are approximated as a set of harmonic
oscillators with frequencies $\omega_j$, $N_v(E)$
can be calculated using
the Beyer-Swinehart algorithm
\citep{Beyer+Swinehart_1973,Stein+Rabinovitch_1973}.
We will consider silicate nanoparticles as an example, but our conclusions
are insensitive to the detailed composition, and similar results would
be obtained for PAHs or for Fe nanoparticles.

The spectrum of vibrational modes for silicates was discussed
by \citet[][hereafter DL01]{Draine+Li_2001}.
DL01 found that the experimental specific heats 
\citep{Leger+Jura+Omont_1985}
for basalt glass
(50\% SiO$_2$, 50\% metal oxides by mass) and obsidian glass
(75\% SiO$_2$, 25\% metal oxides by mass)
could be reproduced if two-thirds of the modes were distributed
according to a 2-dimensional Debye model with Debye temperature
$\Theta_2=500\K$, and one-third of the modes according to a 3-dimensional
Debye model with Debye temperature $\Theta_3=1500\K$.
The lowest-frequency mode is estimated to have
\beq \label{eq:mode1}
\hbar\omega_1=
\frac{k\Theta_2}
{2^{5/3}(N-2)^{2/3}}
~~~.
\eeq
The $(N-2)^{-2/3}$ dependence arises from the assumption that some of
the modes are distributed as for a 2-dimensional Debye model.\footnote{
    For the 3-dimensional Debye model, the lowest frequency
    mode would scale as $\omega_1\propto (N-2)^{-1/3}$.
    }
While surprising, this model does reproduce the
measured specific heat for bulk basalt and obsidian down to $10\K$
\citep[see Fig.\ 2 of][]{Draine+Li_2001}.
For $\Theta_2=500\K$ and $N=40$, Eq.\ (\ref{eq:mode1})
gives $\hbar\omega_1/hc=10\cm^{-1}$.
The lowest-frequency vibration will presumably be a bending or torsional
mode of the nanocluster.

\begin{figure}[ht]
\begin{center}
\includegraphics[angle=0,width=8.0cm,
                 clip=true,trim=0.5cm 5.0cm 0.5cm 2.5cm]
{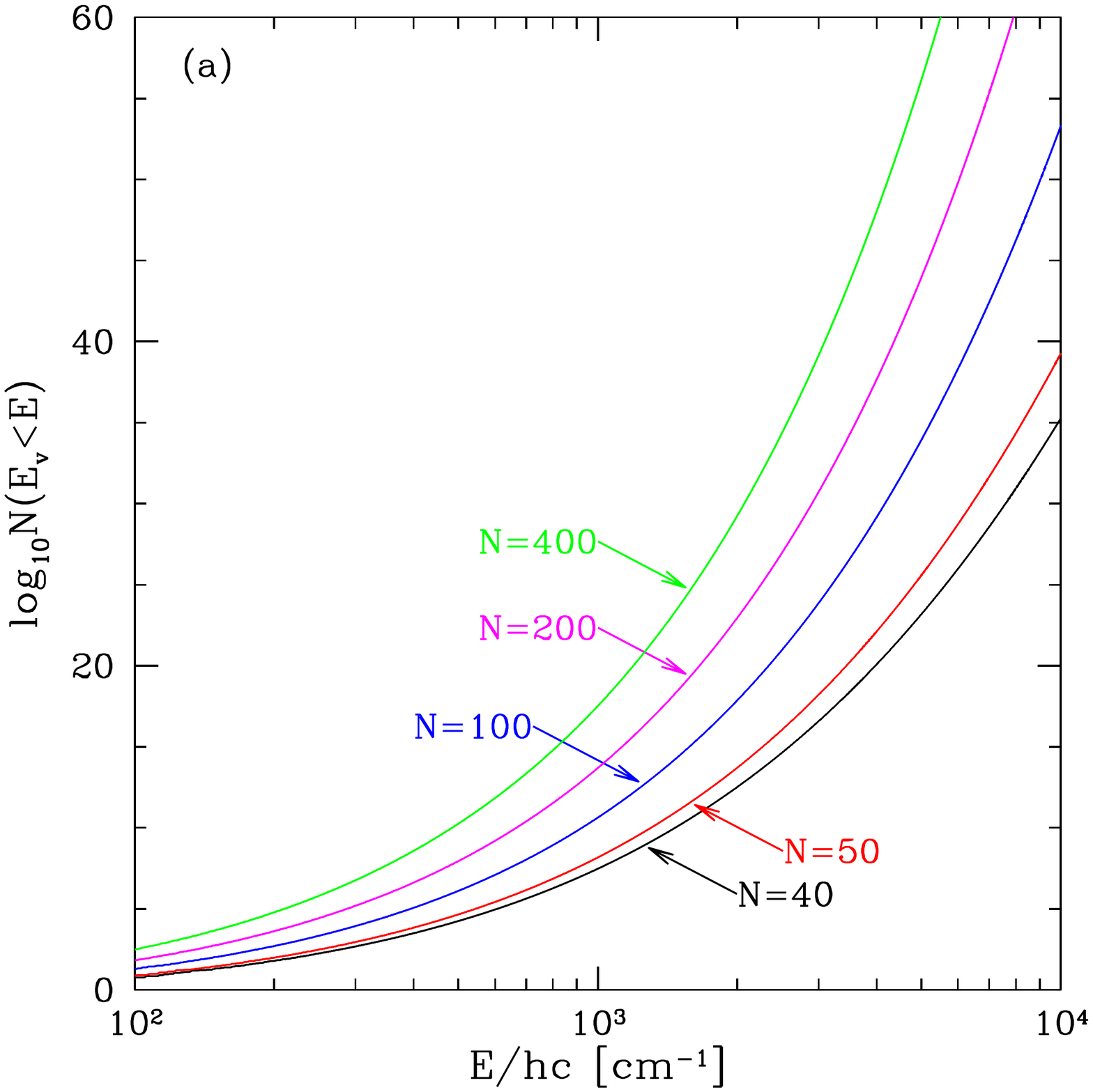}
\includegraphics[angle=0,width=8.0cm,
                 clip=true,trim=0.5cm 5.0cm 0.5cm 2.5cm]
{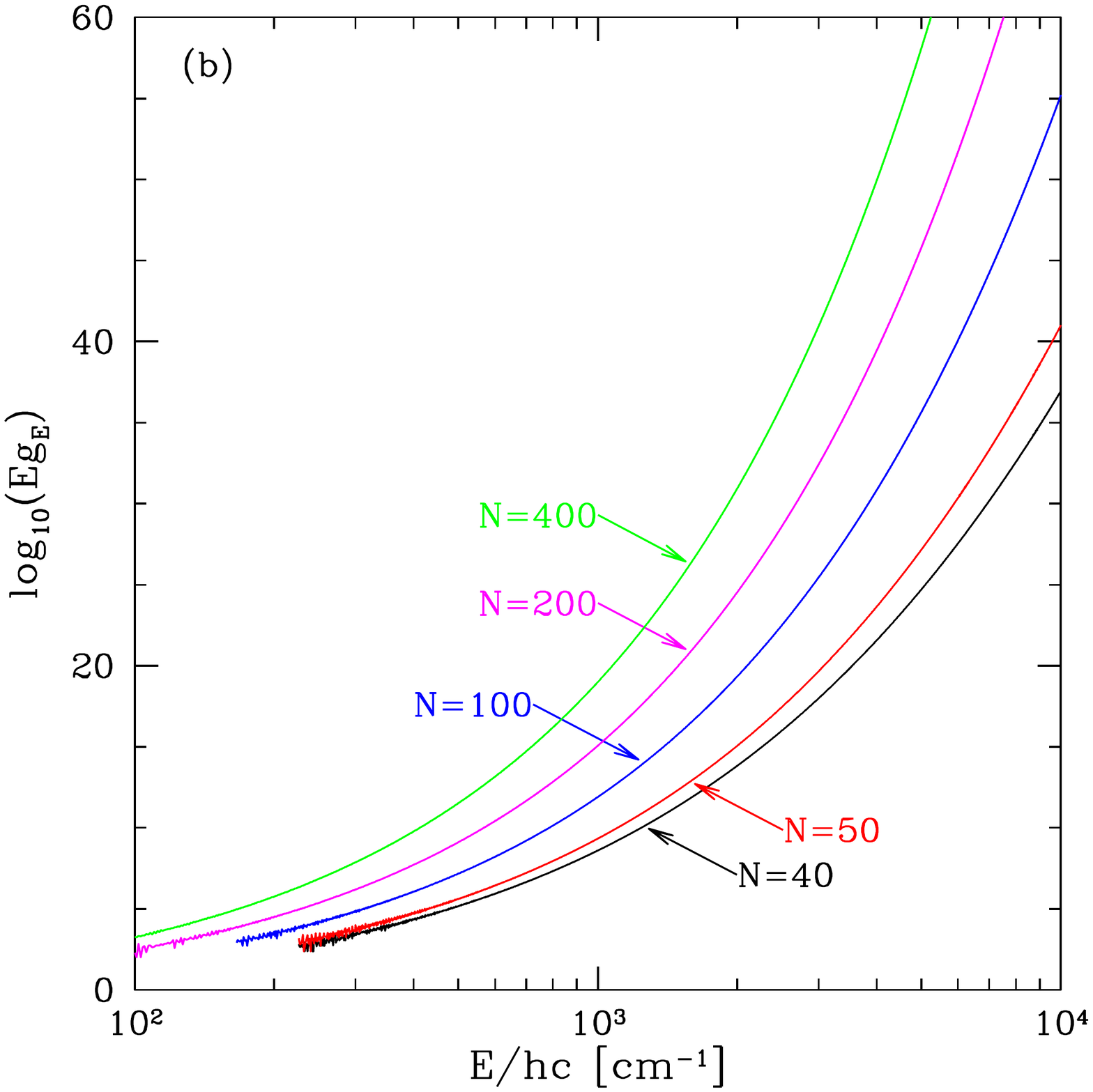}
\caption{\label{fig:NofE}\footnotesize
         (a) The number $N_v(E)$ of distinct vibrational states 
         with vibrational energy $\Evib<E$, for
         nanosilicate clusters with $N=40,50,100,200$ and $400$ atoms.
         $N_v(E)$ is computed using the Beyer-Swinehart
         algorithm and the fundamental
         mode spectrum estimated by \citet{Draine+Li_2001}.
         (b) $E g_E$, where $g_E\equiv dN_v/dE$ is 
         the vibrational density of states.
         }
\end{center}
\end{figure}

Using the mode spectrum prescription from DL01, 
$N_v(E)$ is calculated using the Beyer-Swinehart algorithm.
The result is shown in Figure \ref{fig:NofE}a for 5 values of $N$.
The vibrational density of states $g_E\equiv dN_v/dE$ is shown in 
Figure \ref{fig:NofE}b, where we have averaged over bins of width
$\Delta E/hc=1\cm^{-1}$; the ``noise'' at the lowest energies arises
from the stepwise character of the function $N_v(E)$.

For vibrational energies $\Evib/hc\gtsim 5000\cm^{-1}$, 
the number of states $N_v(E)$ is huge ($>10^{20}$), 
and the vibrational states can
be treated as a continuum, even for a nanoparticle with as few as 40
atoms.  However, for low energies, the
discreteness of the vibrational spectrum can have important consequences.

\section{\label{sec:quantum suppression}
         Quantum Suppression of Internal Relaxation}

\begin{figure}[ht]
\begin{center}
\includegraphics[angle=0,width=8.0cm,
                 clip=true,trim=0.5cm 5.0cm 0.5cm 2.5cm]
{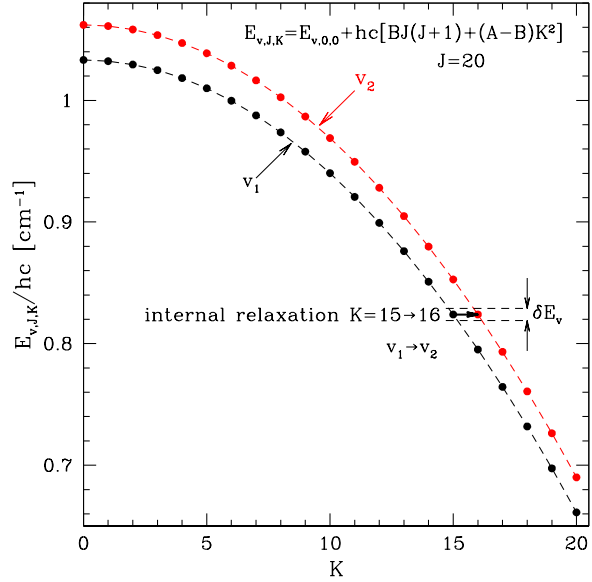}
\caption{\label{fig:Ktransition}\footnotesize
         The energy difference $E_{v,J,K}-E_{v_1,0,0}$
for $J=15$.
The transition $(v_1,J,K)\rightarrow(v_2,J,K+1)$ can take place
only if $E_{v_2,J,K+1}=E_{v_1,J,K}\pm\delta \Evib$.
}
\end{center}
\end{figure}

\begin{figure}[ht]
\begin{center}
\includegraphics[angle=0,width=10.0cm,
                 clip=true,trim=0.5cm 0.5cm 0.5cm 0.5cm]
{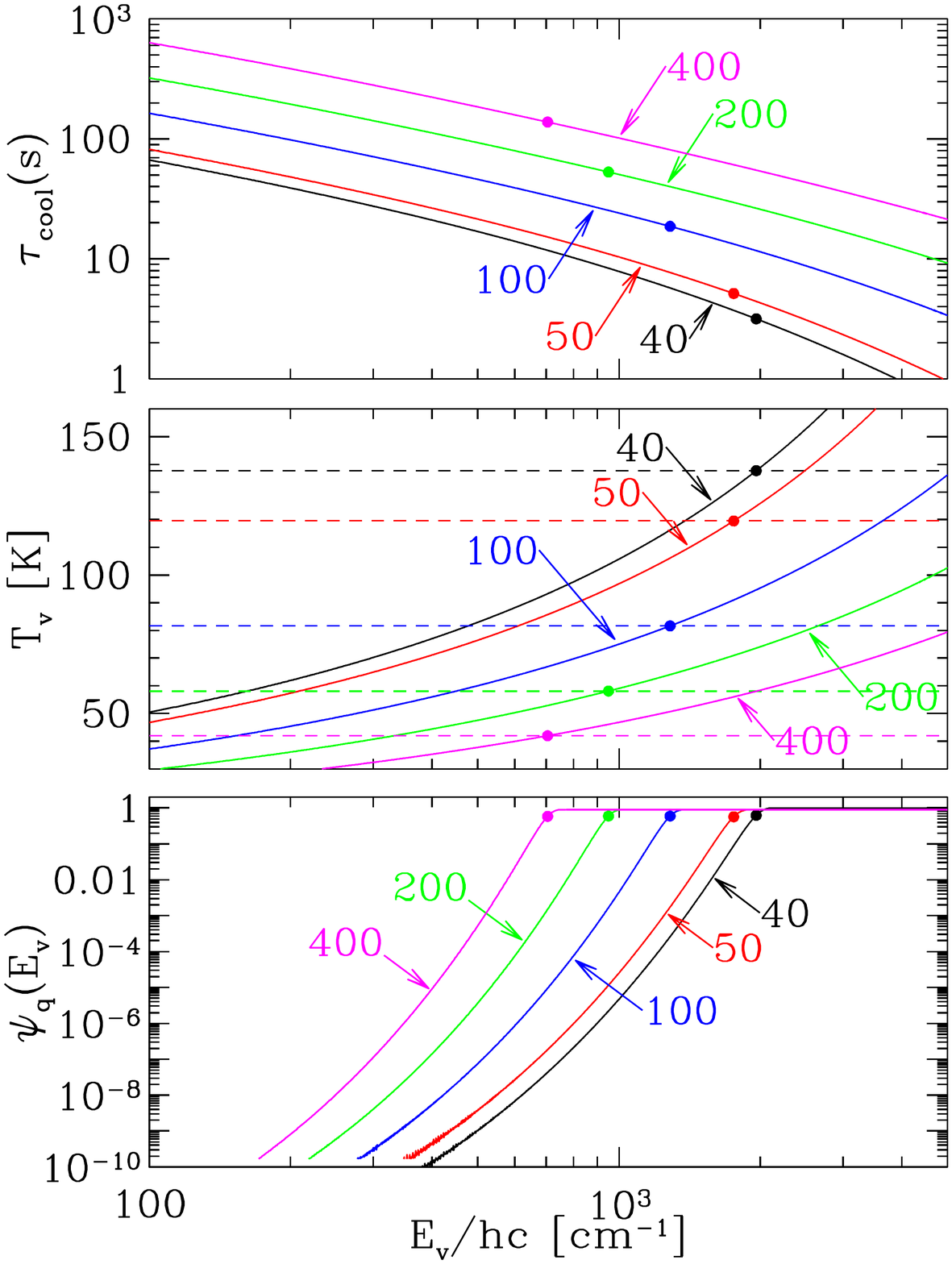}
\caption{\label{fig:tau}\footnotesize
         Cooling time $\tau_{\rm cool}$, vibrational temperature $\Tvib$, 
         and quantum suppression factor 
         $\psi_{\rm q}(\Evib)$, all as functions
         of the vibrational energy $\Evib$, for silicate nanoparticles
         containing $N=40, 50, 100, 200, 400$ atoms.
         For each case, the dot indicates the energy at which
         suppression of dissipation takes effect as the grain cools.
}
\end{center}
\end{figure}

From Fig.\ \ref{fig:Erot} we see that an oblate grain with fixed $J$
can reduce $\Erot$ by increasing $K$, e.g., $K=15\rightarrow16$.
However, energy conservation requires that this energy be transferred
to vibrational modes.
For a transition $K\rightarrow K+1$ to be possible, 
one must have $K<J$ {\it and} there must be {\it another}
vibrational state $v_2$ such that
\beq \label{eq:K transition}
\left| E_{v_2,J,K+1}-E_{v_1,J,K}\right| \ltsim \delta E
~~~,
\eeq
where $\delta E$ is the width of the energy level due to
radiative or collisional broadening.
Figure \ref{fig:Ktransition} shows such a transition.
If we approximate $B_{v_2}\approx B_{v_1}$, $A_{v_2}\approx A_{v_1}$,
then Eq.\ (\ref{eq:K transition}) becomes
\beq \label{eq:Kresonance}
E_{v_2,0,0}=E_{v_1,0,0} + hc(B_v-A_v)(2K+1) \,\pm \delta E
~~~.
\eeq
The probability of (\ref{eq:Kresonance}) being satisfied, i.e., 
for a state $v_2$ to be available at the required energy,
is approximately
\beq \label{eq:psi}
\psi_{\rm q}(\Evib) = 1 - \exp\left(-g_E\,\delta E\right) 
~~~,
\eeq
where $g_E$ is the vibrational density of states.
For a tumbling grain, we will take the rate of vibration-rotation
energy exchange (due to viscoelastic dissipation or other processes)
to be the ``bulk'' rate multiplied by $\psi_{\rm q}(\Evib)$,
which we will refer to as the quantum suppression factor.

Condition (\ref{eq:Kresonance}) applies to viscoelastic
dissipation, which exchanges energy between kinetic energy of
rotation and vibrational energy while the lattice
angular momentum $J$ remains constant.
The processes of ``Barnett relaxation'' \citep{Purcell_1979} -- 
where the rotational
kinetic energy can be reduced if some of the angular momentum
is taken up by the system of electron spins -- 
and ``nuclear spin relaxation'' \citep{Lazarian+Draine_1999b} --
where angular momentum is transferred to the system of nuclear spins --
are slightly different from viscoelastic damping,
because some of the lattice 
angular momentum is transferred to the electron or nuclear spin systems,
and
the lattice angular momentum quantum
number $J\rightarrow J-1$.  Such transitions are discussed in
\S\ref{sec:Jchanging}, but the same quantum suppression factor
$\psi_{\rm q}$ applies.

\begin{figure}[ht]
\begin{center}
\includegraphics[angle=0,width=8.0cm,
                 clip=true,trim=0.5cm 0.5cm 0.5cm 0.5cm]
                {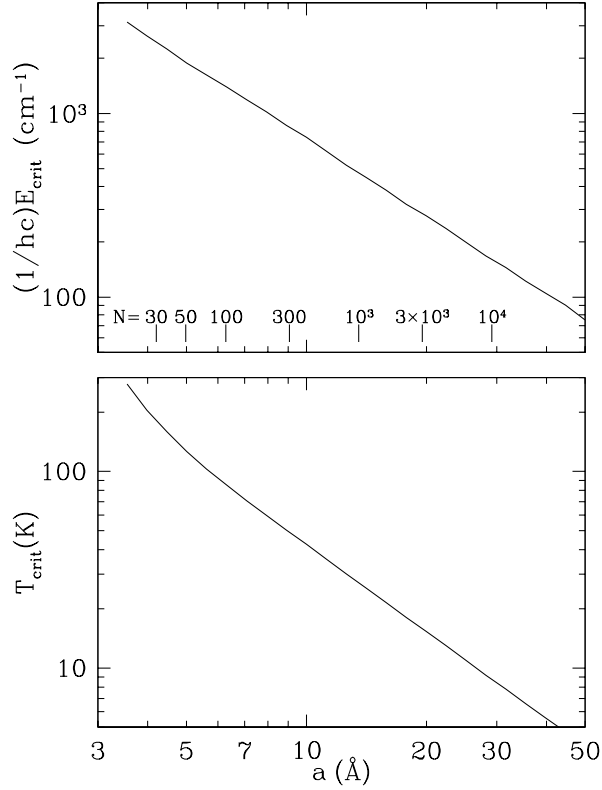}
\caption{\label{fig:etcrit}\footnotesize
         Vibrational energy $E_\crit$ and vibrational
         temperature $T_\crit$ where $g_E\delta E=1$, above which
         the vibrational states can be treated as a continuum.
         When the grain cools below $T_\crit$,
         internal dissipation processes are suppressed.
         }
\end{center}
\end{figure}

The lower panel of Figure \ref{fig:tau} shows 
the quantum suppression factor $\psi_{\rm q}(\Evib)$ 
as a function of vibrational energy $\Evib$ for 5 selected sizes: 
$N=40$, 50, 100, 200, and 400 atoms.
For each case, the dot shows the point where $g_E\delta E = 1$;
for energies below this point, IVRET will be suppressed.
The middle panel shows the vibrational temperature $\Tvib$ as a function of
$E$, and the upper panel shows the cooling time 
$\tau_{\rm cool}\equiv \Evib/|d\Evib/dt|_{\rm rad}$ where
$|d\Evib/dt|_{\rm rad}$ is the thermal power radiated by the grain.
We define the critical temperature $T_\crit$ to be the
vibrational temperature at which $g_E\delta E=1$.
Figure \ref{fig:etcrit} shows the critical energy $E_\crit$ and
critical vibrational temperature $T_\crit$ as a function of
nanoparticle size.

\section{\label{sec:align a with J}
         Alignment of the Grain Body with Angular Momentum $\bJ$}

Consider a grain with angular momentum quantum number $J$.
Alignment of the grain axis $\bahat_1$ with $\bJ$ is measured
by
\beq
\langle\cos^2\theta_{\bahat_1\bJ}\rangle = \frac{\langle K^2\rangle}{J(J+1)}
~~~,
\eeq
where $\theta_{\bahat_1\bJ}$ is the angle between $\bahat_1$ and $\bJ$.
If vibration-rotation energy exchange is rapid, then
the tumbling grain will have fluctuating $K$, with
the probability of being in state $K$ given by
\beq \label{eq:p_K}
p_K(J,\Tvib) = C^{-1}\exp\left[\frac{hc(B_v-A_v)K^2}{k\Tvib(E)}\right] 
\quad\quad K=-J, ..., J
\eeq
where $C=\sum_{K=-J}^J \exp\left[hc(B_v-A_v)K^2/k\Tvib\right]$.
The tumbling grain will have $\langle K^2\rangle = \sum K^2 p_K$.

Single-photon heating of a small nanoparticle will raise it to a high
temperature, resulting in
near-random orientation 
($\langle\cos^2\theta_{\bahat_1\bJ}\rangle\approx 1/3$).
So long as IVRET is rapid,
$\langle\cos^2\theta_{\bahat_1\bJ}\rangle$ will gradually increase 
as the grain cools,
and the states of lower rotational energy are increasingly favored.
When the grain temperature falls to $T_\crit$, the rate of
internal dissipation will be suppressed.  As shown in Figure \ref{fig:tau},
the onset of suppression for a nanoparticle with $N=100$ atoms is
at $\Tvib\approx 75\K$, and by the time the temperature has dropped to
$\sim 65\K$ the suppression factor $\psi_{\rm q}\approx 10^{-6}$.


We have $T_\crit\approx 140\K$ for $N=40$, and $\sim$$40\K$ for
$N=400$.
Because of the rapid drop in $\psi_{\rm q}(\Evib)$ 
when $T$ drops below $T_\crit$,
we will approximate 
vibration-rotation energy exchange as rapid provided
$T> T_\crit$, but negligibly slow when $T<T_\crit$.
Thus, after absorbing a starlight photon that heats it to $T>T_\crit$, 
the nanoparticle will have full internal
relaxation, with $p_K$ given by Eq.\ (\ref{eq:p_K}),
as it cools down until reaching temperature $T_\crit$,
at which time the angle $\theta_{\bahat_1\bJ}$ is frozen until the
next starlight photon heating event, or $\bJ$ is changed by collisions
or radiation.
For a grain undergoing stochastic heating by starlight photons, 
let $p_E$ be the probability of being in energy bin $E$.
Then, for grains with angular momentum quantum number $J$
\beq \label{eq:<cos2theta_aJ>}
\langle\cos^2\theta_{\bahat_1\bJ}\rangle \approx
\frac{1}{J(J+1)}\left[ 
\sum_{E>E_\crit} p_E \sum_K K^2 p_K(J,T_E) + 
\left(\sum_{E=0}^{E_\crit} p_E\right)\sum_K K^2 p_K(J,T_\crit)\right]
~~~.
\eeq
We will evaluate 
$\langle\cos^2\theta_{\bahat_1\bJ}\rangle$ below after
discussion of the excitation of $J$.

\section{\label{sec:Jchanging}
         Quantum Suppression of Magnetic Dissipation and Alignment with
         $\bB_0$}

Now consider a nanoparticle spinning in the static interstellar
magnetic field $\bB_0$.
The unpaired electron spins in the nanoparticle couple
to $\bB_0$.
If $\bJ$ has a component perpendicular to $\bB_0$, then the
(weak) magnetization of the grain will lag (in the grain frame,
there is a rotating component of the magnetic field),  
and the coupling of $\bB_0$ to the unpaired spins will
exert a torque on the spinning grain, acting to reduce the
component of the total angular momentum $\bJ$ that is perpendicular
to $\bB_0$.
This is the \citet{Davis+Greenstein_1951}
mechanism for alignment of $\bJ$ by magnetic dissipation.

Let $\Jpara$ and $\Jperp$ 
be the components of the angular momentum
parallel and perpendicular to $\bB_0$.
In a large paramagnetic grain at temperature $\Tvib$, 
magnetic dissipation will cause $\Jperp$ to
change at a rate
\beqa
\left(\frac{d\Jperp}{dt}\right)_{\rm DG} 
&=& - \frac{\Jperp}{\tau_{\rm DG,0}}
\left(\frac{T_0}{\Tvib}\right)
\\ \label{eq:tauDG0}
\tau_{\rm DG,0} &\equiv& \frac{2\rho a^2}{5 B_0^2 K_0}
~~~,\eeqa
where
\beq \label{eq:K_0 PM}
K_0\approx 10^{-13}\left(\frac{18\K}{T_0}\right)\s 
\eeq
for normal paramagnetic dissipation \citep{Jones+Spitzer_1967}.
For this estimate for $K_0$ to apply to a nanoparticle, 
there should be at least a few
unpaired spins present in the system, so that (1) there are spins to
respond to the magnetic field, and (2) there will be spin-spin coupling
as per the \citet{Jones+Spitzer_1967} estimate for $K_0$.

\begin{figure}[ht]
\begin{center}
\includegraphics[angle=0,width=8.0cm,
                 clip=true,trim=0.5cm 5.0cm 0.5cm 2.5cm]
{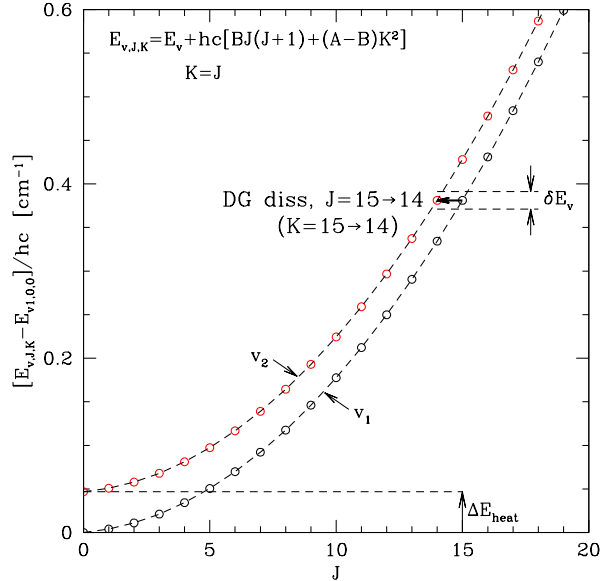}
\caption{\label{fig:DGdiss}\footnotesize
         Transition from magnetic dissipation in a grain
         rotating in a static magnetic field.
         $\Delta J=-1$, and rotational kinetic energy is converted
         to heat $\Delta E_{\rm heat}$.
         Here we show an example where the grain initially had
         $J_1=K_1=15$, and $\Delta K=-1$.
         Levels with $K=J$ are shown for vibrational states $v_1$ and $v_2$.
}
\end{center}
\end{figure}
\begin{figure}[ht]
\begin{center}
\includegraphics[angle=0,width=8.0cm,
                 clip=true,trim=0.5cm 5.0cm 0.5cm 2.5cm]
                {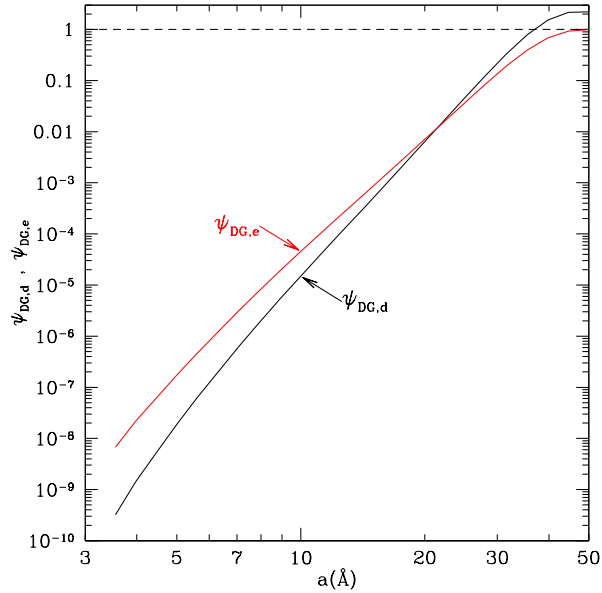}
\caption{\label{fig:psi_dg}\footnotesize
         Quantum suppression factor $\psi_{\rm DG,d}$ for
         Davis-Greenstein paramagnetic alignment, as a function of
         radius for nanosilicate grains
         stochastically heated by the \citet{Mathis+Mezger+Panagia_1983}
         interstellar radiation field.
         The related suppression factor $\psi_{\rm DG,e}$
         for excitation of rotation perpendicular to $\bB_0$
         is also shown.
         }
\end{center}
\end{figure}
\begin{figure}[ht]
\begin{center}
\includegraphics[angle=270,width=8.0cm,
                 clip=true,trim=0.5cm 0.5cm 0.5cm 0.5cm]
{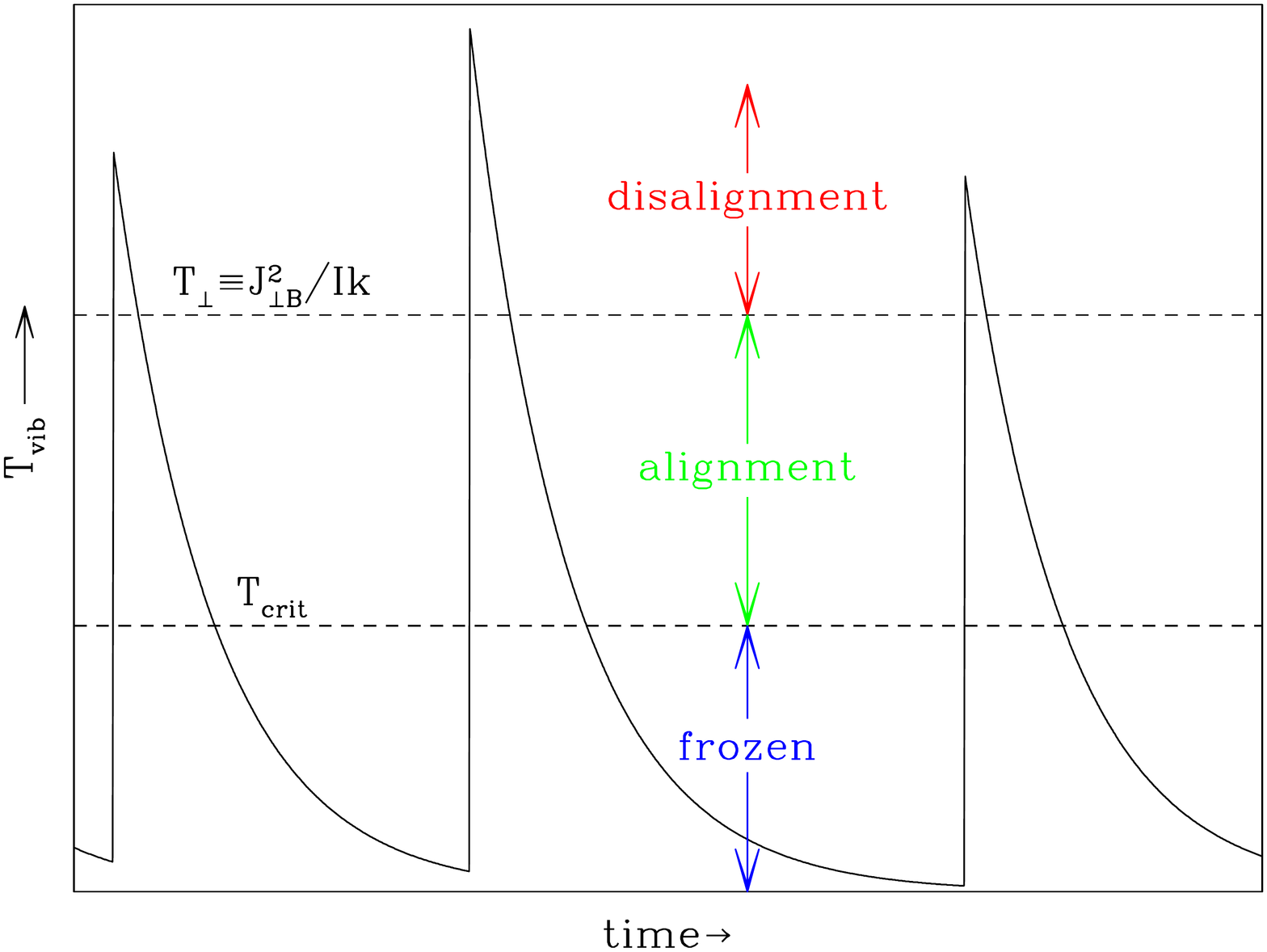}
\caption{\label{fig:schema}\footnotesize
         Schematic of vibrational temperature vs.\ time
         for a grain undergoing stochastic heating and radiative
         cooling.
         When the grain vibrational temperature 
         $\Tvib>T_\perp\equiv\Jperp^2/Ik\approx T_{\rm gas}$,
         the effect of the static magnetic field $B_0$ is to
         increase $\Jperp^2$, acting to disalign $\bJ$ with $\bB_0$.
         When $\Tvib<T_\perp$, paramagnetic dissipation tends to
         {\it reduce} $\Jperp^2$, acting to align $\bJ$ with $\bB_0$.
         When $\Tvib$ falls below $T_\crit$, dissipation is
         suppressed, and the alignment is effectively frozen until the next
         heating event.
         }
\end{center}
\end{figure}

In a single-domain ferromagnetic or ferrimagnetic grain, the
spins will be spontaneously aligned, and the magnetization dynamics
are quite different from the paramagnetic case.
Dissipation in ferromagnetic materials at high frequencies
has been discussed by \citet{Draine+Hensley_2013}, with attention
to Davis-Greenstein alignment.
Using the Gilbert equation \citep{Gilbert_2004}
for the dynamical magnetization
with Gilbert parameter $\alpha_G\approx0.2$,
$K_0$ for a pure Fe grain is estimated to be 
\citep[see][eq.\ 97]{Draine+Hensley_2013}
\beq \label{eq:K_0 FM}
K_0\approx 3\times10^{-13}(1+\cos^2\Theta)\s
~~~,
\eeq
where $\Theta$ is the angle between the spontaneous
magnetization $\bM$ and $\bJ$.
For paramagnetism, $K_0\propto 1/T_0$, but 
for ferromagnetism, $K_0$ does not depend on grain temperature
(provided the temperature is well below the Curie temperature
$T_{\rm C}\approx1100\K$).
The estimate for $K_0$ for metallic Fe is (coincidentally)
only a factor of a few larger
than the classical estimate
for paramagnetism (\ref{eq:K_0 PM}) at the typical grain
temperature $T_0\approx18\K$.
The dependence on $\Theta$ causes
ferromagnetic dissipation to be faster if the spontaneous magnetization
direction is close to $\bJ$, but $(1+\cos^2\Theta)$ is at most
a factor of two.

The time-dependent torques experienced by the unpaired spins
are transferred to the lattice, which will both excite
lattice vibrations (heat) and reduce the lattice angular momentum.
Energy is conserved: the decrease in rotational kinetic energy
is accompanied by heating of the lattice.
Figure \ref{fig:DGdiss} shows an example of such a transition
$J\rightarrow J-1$.  The example shown has $K\rightarrow K-1$,
but other values of $\Delta K$ can also take place.

Because energy must be conserved, magnetic dissipation
can only take place if there is a suitable energy level $v_2$ such
that (see Fig.\ \ref{fig:DGdiss})
\beq \label{eq:DGheatres}
E_{v_2,0,0}=E_{v_1,0,0} + \Delta E_{\rm heat} \,\pm \delta E
~~~,
\eeq
where $\delta E$ is the ``level width'', and
\beq
\Delta E_{\rm heat} = 2B_vJ + (A_v-B_v)(K_1^2-K_2^2)
~~~.
\eeq
The likelihood of a vibrational state $v_2$ being available so that
(\ref{eq:DGheatres}) can be satisfied is again given by the
function $\psi_{\rm q}(\Evib)$ defined in Eq.\ (\ref{eq:psi}).

For a nanoparticle
heated by starlight photons, $\Tvib$ is a stochastic
function of time, shown schematically in Figure \ref{fig:schema}.
When the grain is vibrationally ``hot'' immediately following
a photon absorption,
the torques due to the
static magnetic field act to {\it disalign} $\bJ$ and $\bB_0$
(by acting to increase $\Jperp^2$)
but when $\Tvib$
drops below $T_\perp\equiv\Jperp^2/Ik$, 
the dissipative torques have a net
{\it aligning} effect (by acting to decrease $\Jperp^2$).
When the temperature falls below $T_\crit$, the dissipation is
strongly suppressed; the suppression factor $\psi_{\rm q}$ falls off
so rapidly that the magnetic torques effectively cease, and
we can think of the alignment as frozen until the next starlight
photon is absorbed.

Averaging over the temperature fluctuations, we take the systematic
aligning torque due to
magnetic dissipation to be
\beqa
\left(\frac{d\Jperp}{dt}\right)_{\rm DG,d} &=& 
- \frac{\Jperp}{\tau_{\rm DG,0}}\psi_{\rm DG,d}
\\
\psi_{\rm DG,d}
&\equiv&
\sum_E p_E \left(\frac{\Tvib(E)}{T_0}\right)^n\times \psi_{\rm q}(E)
~~~.
\eeqa
where $n=-1$ for paramagnetism, and $n=0$ for ferromagnetism or
ferrimagnetism.

The fluctuation-dissipation theorem implies that there must also be
excitation if the lattice temperature $\Tvib>0$:
\beqa
\left\langle\frac{d\Jperp^2}{dt}\right\rangle_{\rm DG,e}
&=&
\frac{4I kT_0}{\tau_{\rm DG,0}}
\psi_{\rm DG,e}
\\
\psi_{\rm DG,e} &\equiv& \sum_E p_E \left(\frac{\Tvib}{T_0}\right)^{n+1}
\psi_{\rm q}(E)
~~~.
\eeqa
where again $n=-1$ for paramagnetism, and $n=0$ for ferromagnetism.
It is this excitation that leads to {\it disalignment} of $\bJ$ and
$\bB_0$.
Figure \ref{fig:psi_dg} shows the suppression factors $\psi_{\rm DG,d}$
and $\psi_{\rm DG,e}$.
The energy distribution functions $p_E$ 
for silicate nanoparticles with radii $a$ heated
by the interstellar radiation field estimated for the
solar neighborhood by \citet{Mathis+Mezger+Panagia_1983}
were calculated following \citet{Draine+Li_2001}.

\citet{Papoular_2016} recently proposed that interstellar grains
can be aligned with the magnetic field $\bB_0$ even if the grain
material has zero magnetic susceptibility, because the ions and
electrons in the grain will experience time-varying Lorentz forces
unless the grain angular velocity $\bomega\parallel\bB_0$.
The mechanism proposed by Papoular 
also relies on dissipation of rotational kinetic energy and
transfer of energy between rotation and vibration, and would
be subject to the same quantum suppression factor that would
apply to the paramagnetic dissipation envisaged by 
\citet{Davis+Greenstein_1951}, or dissipation in superparamagnetic
or ferromagnetic materials.

\section{\label{sec:excitation and alignment}
         Excitation of $J$, and Alignment of $\bJ$ with $\bB$}

We now consider the balance between excitation and damping by the
various torques acting on a spinning grain.
Let $\Jpara$ and $\Jperp$ be the components of the grain
angular momentum parallel and perpendicular to $\bB$.
For an individual grain, $\Jpara(t)$ and $\Jperp(t)$ are
stochastic variables.  For an ensemble, we write
\beqa \label{eq:dJpara2/dt}
\frac{d}{dt} \frac{\langle \Jpara^2\rangle }{2I}
&\approx& 
\frac{G}{\tau_\Ha}\frac{kT_\gas}{2} - 
\frac{F}{\tau_\Ha} \frac{\langle \Jpara^2\rangle}{2I}
- \frac{4\mu_\perp^2}{9c^3}\langle\omega^4\rangle
\frac{\langle \Jpara^2\rangle}{\langle J^2\rangle}
\\ \label{eq:dJperp2/dt}
\frac{d}{dt} \frac{\langle \Jperp^2\rangle }{2I}
&\approx& 
\frac{G}{\tau_\Ha}
kT_\gas - 
\frac{F}{\tau_\Ha} \frac{\langle \Jperp^2\rangle}{2I}
- \frac{4\mu_\perp^2}{9c^3}\langle\omega^4\rangle
\frac{\langle \Jperp^2\rangle}{\langle J^2\rangle}
-\frac{(\langle \Jperp^2\rangle-J_0^2)\psi_{\rm DG,d}}
{I\tau_{\rm DG,0}}
~~~.
\eeqa
Here we approximate the grains as spherical, so that we need
not consider the orientation of the body of the grain relative
to $\bJ$.
The characteristic timescale  
\beq \label{eq:tauH}
\tau_\Ha \equiv \frac{4\rho a}{\nH\mH}
\left(\frac{\pi \mH}{8kT_\gas}\right)^{1/2}
\approx 2\times10^{11}\left(\frac{30\cm^{-3}}{\nH}\right)
\left(\frac{a}{10^{-7}\cm}\right)
\left(\frac{100\K}{T_\gas}\right)^{1/2}\s
\eeq
is the rotational damping time for a neutral grain in a gas of atomic H.
$F(a)$ and $G(a)$ are 
dimensionless factors introduced by \citet{Draine+Lazarian_1998b}, allowing
for the actual
rates for rotational damping and excitation arising from
partial ionization of the gas, charging of the grain, and
the effects of starlight and infrared emission from the grain.
Note that $\dot{N}_\abs\tau_\Ha\gg 1$ (see Eq.\ \ref{eq:Ndotabs}): 
temperature fluctuations
due to stochastic heating occur on a time much shorter than
the time for angular momentum variations, alignment, etc.

Loss of rotational kinetic energy from pure-rotational electric
dipole radiation varies as $\omega^4$, but also depends on the
orientation of the electric dipole moment with respect to the
grain's rotation axis.  Here we take $\mu_\perp^2$ to be
an appropriately-averaged mean square dipole moment perpendicular
to the rotation axis.
Following \citet{Draine+Lazarian_1998b}, we suppose that
\beq
\mu_\perp \approx \beta_0 \sqrt{N}
~~~,
\eeq
where $N$ is the number of atoms in the nanoparticle and
$\beta_0$ is a constant.

$F$ and $G$ are functions of the grain charge state and radius, as well
as the ionization and temperature of the gas, and have
been estimated for a variety of environments
\citep{Draine+Lazarian_1998b,Ali-Haimoud+Hirata+Dickinson_2009,
Hoang+Draine+Lazarian_2010}.
For the present illustration we take $F(a)$ and $G(a)$ calculated by
\citet{Hensley+Draine_2016a} for silicate particles with
electric dipole moments corresponding to $\beta_0=0.3{\,\rm D}$
and cold neutral medium (CNM) conditions (see Table \ref{tab:params}).

$\tau_{\rm DG,0}$ is the ``classical'' 
Davis-Greenstein alignment time if the grain temperature were $T_0$.
The factors $\psi_{\rm DG,d}$ and $\psi_{\rm DG,e}$ include the effects
of variations in temperature away from the nominal temperature
$T_0$ as well as the quantum suppression of fluctuations at temperatures
below $T_\crit$.
$J_0^2$ is the mean value of $J_\perp^2$ if the only 
torques were from paramagnetic dissipation and the associated thermal
fluctuations:
\beq
\frac{J_0^2}{2I} = 
\frac{\sum_E P_E \psi_q(E) \tau_{DG}^{-1} kT_E}
{\sum_E P_E \psi_q(E) \tau_{DG}^{-1}}
= kT_0 \frac{\psi_{\rm DG,e}}{\psi_{\rm DG,d}}
~~~.
\eeq 
We average over $E$ because 
the grain undergoes thermal fluctuations on a time short compared
to the characteristic rotational damping time $\tau_\Ha/F$.

\begin{figure}[ht]
\begin{center}
\includegraphics[angle=0,width=8.0cm,
                 clip=true,trim=0.5cm 0.5cm 0.5cm 0.5cm]
{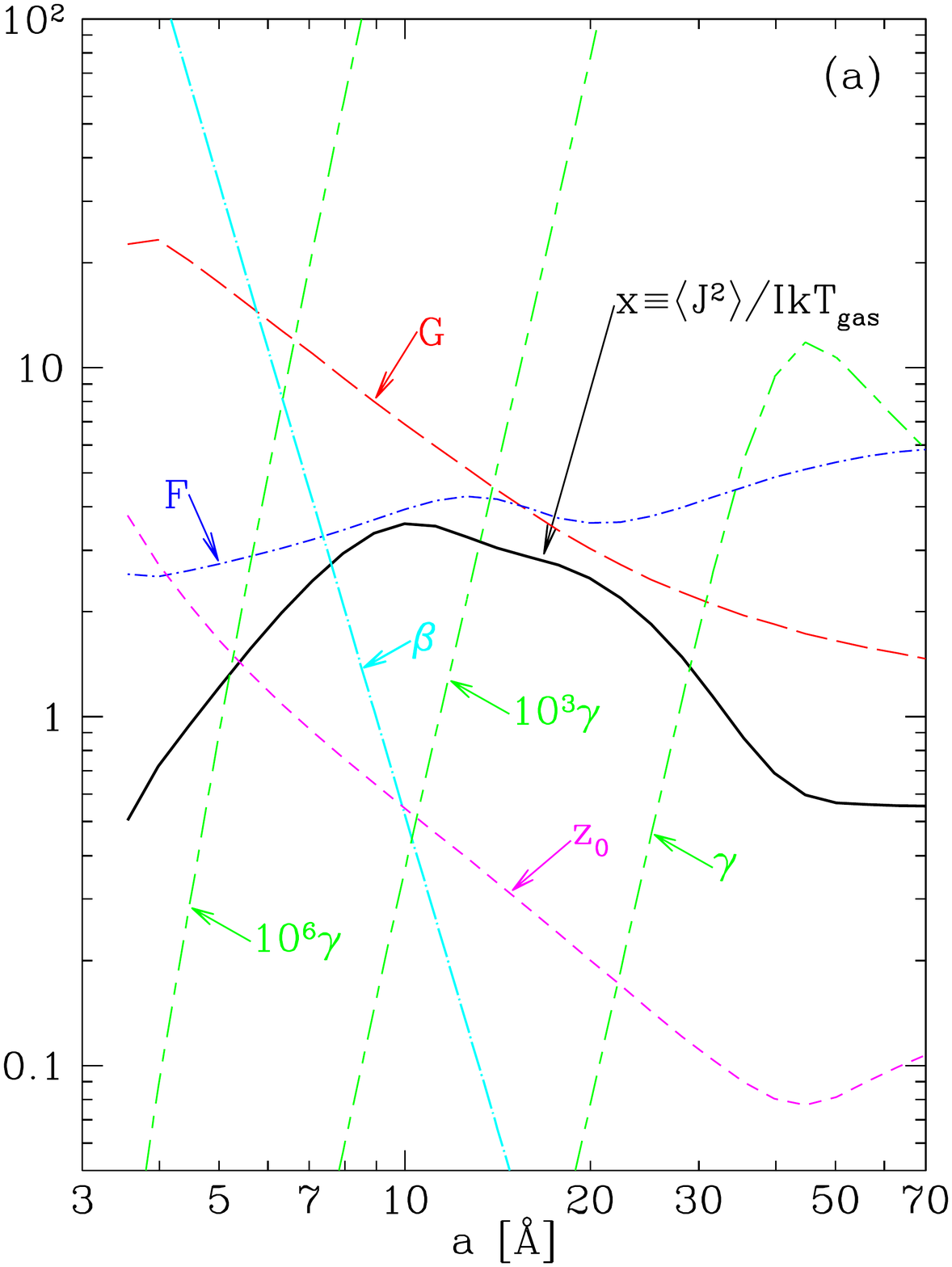}
\includegraphics[angle=0,width=8.0cm,
                 clip=true,trim=0.5cm 0.5cm 0.5cm 0.5cm]
{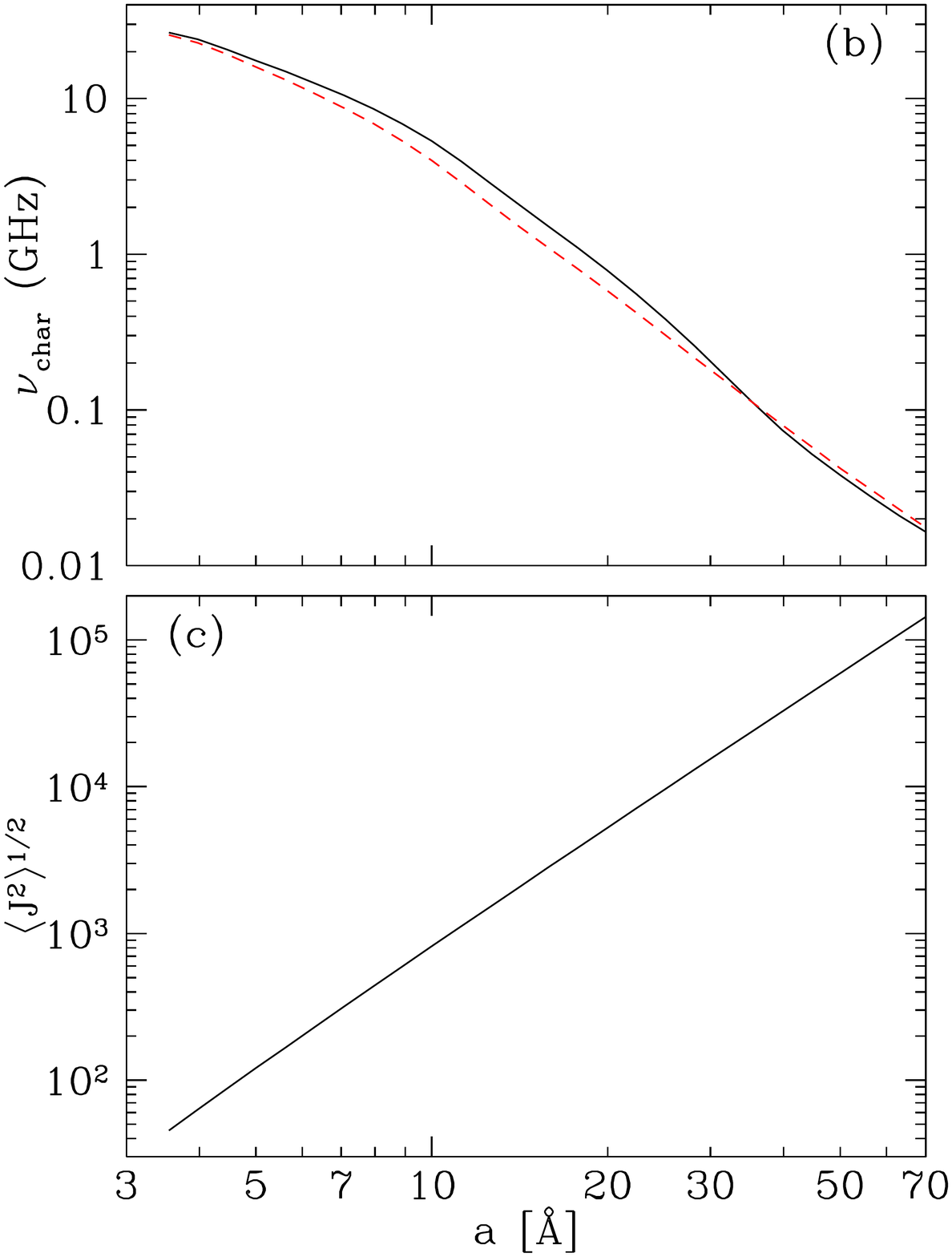}
\caption{\label{fig:frot}\footnotesize
     (a) The dimensionless damping and excitation factors $F$ and $G$,
     the dimensionless parameters 
     characterizing electric dipole damping ($\beta$), 
     magnetic dissipation ($\gamma$),
     and magnetic excitation ($z_0$),
     and the solution $x$ as a function of grain size $a$.
     (b) The characteristic emission frequency as a function of grain size.
     Solid line: including quantum suppression of magnetic dissipation
     and excitation.
     Dashed: neglecting quantum suppression.
     (c) Characteristic angular momentum quantum number $J$ vs.\ $a$.
     }
\end{center}
\end{figure}

We introduce dimensionless parameters
\beqa
\beta &\equiv& \frac{8\mu_\perp^2}{9c^3}
\frac{\langle\omega^4\rangle}{\langle \omega^2\rangle^2}
\frac{kT_\gas}{I^2} \tau_\Ha
\\
\gamma &\equiv& \frac{2\psi_{\rm DG,d}\tau_\Ha}{\tau_{\rm DG,0}}
\\
z_0 &=& \frac{J_0^2}{2IkT_\gas} = 
\frac{T_0}{T_\gas}\frac{\psi_{\rm DG,e}}{\psi_{\rm DG,d}}
~~~,
\eeqa
and dimensionless variables
\beqa
x &\equiv& \frac{\langle \Jperp^2\rangle+\langle\Jpara^2\rangle}{IkT_\gas}
\\
y &\equiv& 
\frac{\langle \Jperp^2\rangle}{\langle \Jperp^2\rangle+\langle\Jpara^2\rangle}
~~~.
\eeqa
Figure \ref{fig:frot}a shows $F$, $G$, $\beta$, $\gamma$, and $z_0$
as functions of grain size $a$ for CNM conditions.

\begin{table}[hb]
{\footnotesize
\begin{center}
\caption{\label{tab:params}
         Environmental Conditions and Grain Properties}
\begin{tabular}{ccc}
\hline
H nucleon density & $\nH$ & $30\cm^{-3}$ \\
electron density & $n_e$ & $0.03\cm^{-3}$ \\
gas temperature & $T_{\rm gas}$   & $100\K$ \\
magnetic field & $B_0$ & $5\muG$ \\
starlight & \multicolumn{2}{c}{\citet{Mathis+Mezger+Panagia_1983}}\\
silicate density & $\rho$ & $3.4\gm\cm^{-3}$\\
silicate mass/atom & $21.6$\,amu \\
electric dipole param. & $\beta_0$ & 0.3\,D \\
\hline
\end{tabular}
\end{center}
}
\end{table}

The total rotational kinetic energy of the grain is measured
by the ``thermality'' variable $x$ -- 
the ratio of the average grain rotational kinetic
energy to what it would be in LTE with the gas temperature -- 
while $y$ measures the disalignment of $\bJ$ from $\bB_0$.
In statistical steady-state, Eq.\ (\ref{eq:dJpara2/dt}, \ref{eq:dJperp2/dt})
become
\beqa \label{eq:steadystate1}
 \tau_\Ha \frac{d}{dt} x(1-y) = \!\!&0&\!\! = G - F x(1-y) 
- \beta x^2 (1-y)
\\ \label{eq:steadystate2}
\tau_\Ha \frac{d}{dt} xy = \!\!&0&\!\! =
2G - Fxy
- \beta x^2 y - \gamma(xy-z_0)
~~~.
\eeqa

\begin{figure}[ht]
\begin{center}
\includegraphics[angle=0,width=8.0cm,
                 clip=true,trim=0.5cm 5.0cm 0.5cm 2.5cm]
{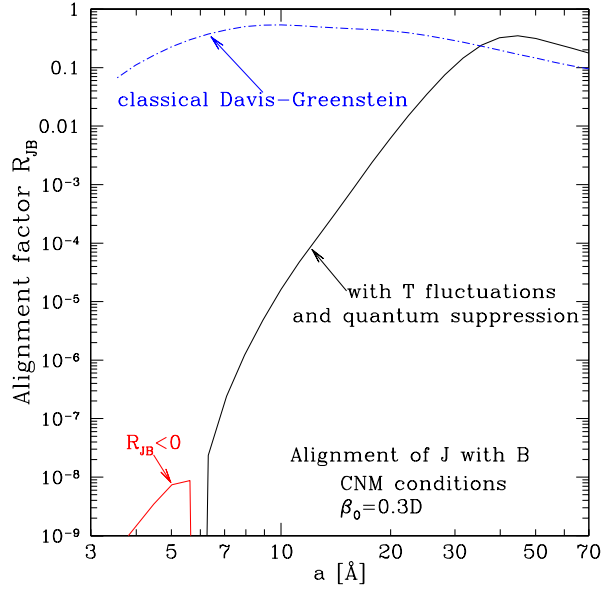}
\caption{\label{fig:rjb}\footnotesize
     Alignment factor $R_{\bJ\bB}$ for spinning grains,
     as a function of grain size $a$.
     Rotational emission has polarization $P<R_{\bJ\bB}$.
     Solid lines: Including effects of
     temperature fluctuations and quantum suppression of dissipation.
     Dot-dashed lines: ``classical'' treatment, neglecting $T$
     fluctuations and assuming a continuous density of states.
     }
\end{center}
\end{figure}

Equations (\ref{eq:steadystate1},\ref{eq:steadystate2}) are solved
to find the steady-state values of $x$ and $y$ for each grain size.
For an assumed $y$ we have
\beq
x= 
\frac{\left[(F+\gamma y)^2 + 4\beta(3G+\gamma z_0)\right]^{1/2}-(F+\gamma y)}
{2\beta}
~~~,
\eeq
and for given $x$, we have
\beq
y = \frac{2}{3} - \gamma\frac{(2/3)x-z_0}{(3F+\gamma)x+3\beta x^2}
~~~.
\eeq
It is straightforward to iterate to find self-consistent $x$ and $y$.
Figure \ref{fig:frot}a shows $x(a)$ for CNM conditions.
The rms rotation frequency for the ensemble is
\beq
\nu_{\rm rms} = \frac{1}{2\pi}\left(\frac{xkT_\gas}{I}\right)^{1/2}
= \frac{1}{2\pi}
\left(\frac{15}{8\pi} \frac{xkT_{\rm gas}}{\rho a^5}\right)^{1/2}
\eeq
and the characteristic emission frequency is\footnote{
Because the radiated power $\propto\omega^4$, we take
$\omega_{\rm char}=\langle\omega^4\rangle^{1/4}$.
For a thermal distribution, $\langle\omega^4\rangle^{1/4}=(5/3)^{1/4}\langle
\omega^2\rangle^{1/2}$.}
\beq
\nu_{\rm char}= \frac{1}{2\pi}
\langle\omega^4\rangle^{1/4}
\approx \left(\frac{5}{3}\right)^{1/4}\nu_{\rm rms}
~~~.
\eeq
Figure \ref{fig:frot}b shows 
$\nu_{\rm char}$ 
for spinning grains as a function of radius $a$, for CNM conditions
and $\beta_0=0.3{\,\rm D}$.
We see that only the smallest grains ($a\ltsim8\Angstrom$) have
characteristic rotation frequencies above $10\GHz$, and thus only
the smallest particles contribute significantly to the AME, which
typically peaks in the $\sim$$20-30\GHz$ range
\citep{Planck_AME_2011,Planck_2015_X}.
Figure \ref{fig:frot}c shows the rms rotational quantum number as
a function of $a$.  Even for the smallest size, we have
$\langle J^2\rangle\gg 1$, justifying the classical treament of
the rotational dynamics in Eq.\ (\ref{eq:dJpara2/dt}, \ref{eq:dJperp2/dt}).

\subsection{Alignment of $\bJ$ with $\bB_0$}

Let $\theta_{\bJ\bB}$ be the angle between $\bJ$ and $\bB_0$.
The ensemble has
\beq
\langle\cos^2\theta_{\bJ\bB}\rangle
= 1-y = \frac{1}{3} + \gamma\frac{(2/3)x-z_0}{(3F+\gamma)x+3\beta x^2}
~~~.
\eeq
We define the alignment factor
\beq
R_{\bJ\bB}\equiv 
\frac{3}{2}\left(\langle\cos^2\theta_{\bJ\bB}\rangle-\frac{1}{3}
\right) = \gamma\frac{x-(3/2)z_0}{(3F+\gamma)x+3\beta x^2}
~~~,
\eeq
which varies from 0 to 1 as $\langle\cos^2\theta_{\bJ\bB}\rangle$
varies from 1/3 (random orientations) to 1 (perfect alignment of $\bJ$
with $\bB_0$).

For $\gamma\ll 1$ we see that grain alignment is very small:
$R_{\bJ\bB} \propto \gamma \ll 1$.
Note that $R_{\bJ\bB}$ can
be negative 
when $z_0>(2/3)x$.  Quantum effects suppress magnetic dissipation
when the grain is cold, but not during the brief intervals (following
starlight heating) when it is hot.  As a result, thermal fluctuations
acting to {\it increase} $\langle \Jperp^2\rangle$ 
can be more important than
dissipation, and cause $R_{\bJ\bB}$ to go negative.
However, this only occurs under conditions where quantum suppression
is so effective that $\gamma\ll 1$, and $|R_{\bJ\bB}|\ll 1$. 

Figure \ref{fig:rjb} shows $R_{\bJ\bB}$ as a function of grain size,
and as a function of the characteristic emission frequency $\nu_{\rm char}$.
The solid lines show results where the alignment is calculated including
the quantum suppression of paramagnetic dissipation in small grains
when $T<T_\crit$.  

For comparison, the alignment of $\bJ$ with $\bB_0$
is also calculated assuming
``classical'' paramagnetic dissipation, as in the standard Davis-Greenstein
treatment.  These results are obtained by solving the
same equations (\ref{eq:steadystate1},\ref{eq:steadystate2}) but
setting $\psi_{\rm DG,d}=\psi_{\rm DG,e}=1$ when
evaluating $\gamma$.
The classical Davis-Greenstein treatment predicts 
$R_{\bJ\bB} \approx 3\%$ for grains spinning at $\sim$$30\GHz$, 
whereas when quantum suppression effects
are included, the alignment factor
$R_{\bJ\bB}$ drops to $\sim-10^{-9}$.

Recent calculations of polarization from spinning silicate nanoparticles
\citep{Hoang+Vinh+Lan_2016} and magnetic Fe nanoparticles 
\citep{Hoang+Lazarian_2016a} concluded that magnetic dissipation
processes would be effective at aligning the particles, with
polarization at 30 GHz predicted to be as large as
$\sim$30\% for silicate nanoparticles, and $\sim$40--50\% for Fe
nanoparticles.
However, when the quantum suppression effects considered here are
included, we predict minimal alignment of such particles, with
extremely low polarization above $\sim$10\,GHz.

For radii $a\gtsim50\Angstrom$, the quantum suppression effects
become unimportant (i.e., $\psi_{\rm q}\approx 1$) and the
present treatment coincides with classical Davis-Greenstein alignment
of nanoparticles, aside from the use of fluctuating grain temperatures
rather than assuming a steady temperature $T_0=18\K$.
Figure 10 shows the alignment factor
$R_{\bJ\bB}$ to be decreasing with increasing
grain size for $a\gtsim50\Angstrom$ as the Davis-Greenstein
alignment time $\tau_{\rm DG,0} \propto a^2$ (see Eq.\ \ref{eq:tauDG0})
becomes long compared to the rotational damping time
$\tau_{\rm H}\propto a$ (see Eq.\ \ref{eq:tauH}).
The observed substantial alignment of the larger
``classical'' grains with $a\gtsim 0.1\micron$
is due to the effects of systematic torques that drive
suprathermal rotation \citep{Purcell_1975,Purcell_1979,Lazarian+Draine_1997}
including the important effects of starlight torques
that can both drive $a\gtsim0.1\micron$ grains to suprathermal rotation
\citep{Draine+Weingartner_1996} as well as directly bring the grain
angular momentum into
alignment with $\bB_0$ 
\citep{Draine+Weingartner_1997,
Weingartner+Draine_2003,
Hoang+Lazarian_2009a,
Hoang+Lazarian_2009b,
Lazarian+Hoang_2011,
Hoang+Lazarian_2016b}.
It is also possible that the larger grains may contain
superparamagnetic inclusions that enhance alignment
\citep{Jones+Spitzer_1967,Mathis_1986,Goodman+Whittet_1995}.
The radiative torques that are important for $a\gtsim0.1\micron$ grains
are negligible for the $a\ltsim0.01\micron$ nanoparticles considered
here, and other possible systematic torques due, e.g., to
formation of H$_2$ and photoelectric emission are suppressed by
the ``thermal flipping'' phenomenon \citep{Lazarian+Draine_1999a}
and can be neglected for the nanoparticles discussed here.

\subsection{Alignment of $\bahat_1$ with $\bJ$}

The polarization of microwave emission depends on the orientation
of the grain's angular velocity $\bomega$, which will not
be parallel to $\bJ$ unless $\bJ$ is parallel to the grain's
principal axis.
The alignment of $\bahat_1$ with $\bJ$ is measured by
\beq
R_{\bahat_1\bJ} \equiv \frac{3}{2}
\left(\langle\cos^2\theta_{\bahat_1\bJ}\rangle - \frac{1}{3}\right)
~~~,
\eeq
which again varies between 0 and 1 as the alignment of $\bahat_1$ with
$\bJ$ goes from random
to perfect.

\begin{figure}[ht]
\begin{center}
\includegraphics[angle=0,width=8.0cm,
                 clip=true,trim=0.5cm 5.0cm 0.5cm 2.5cm]
{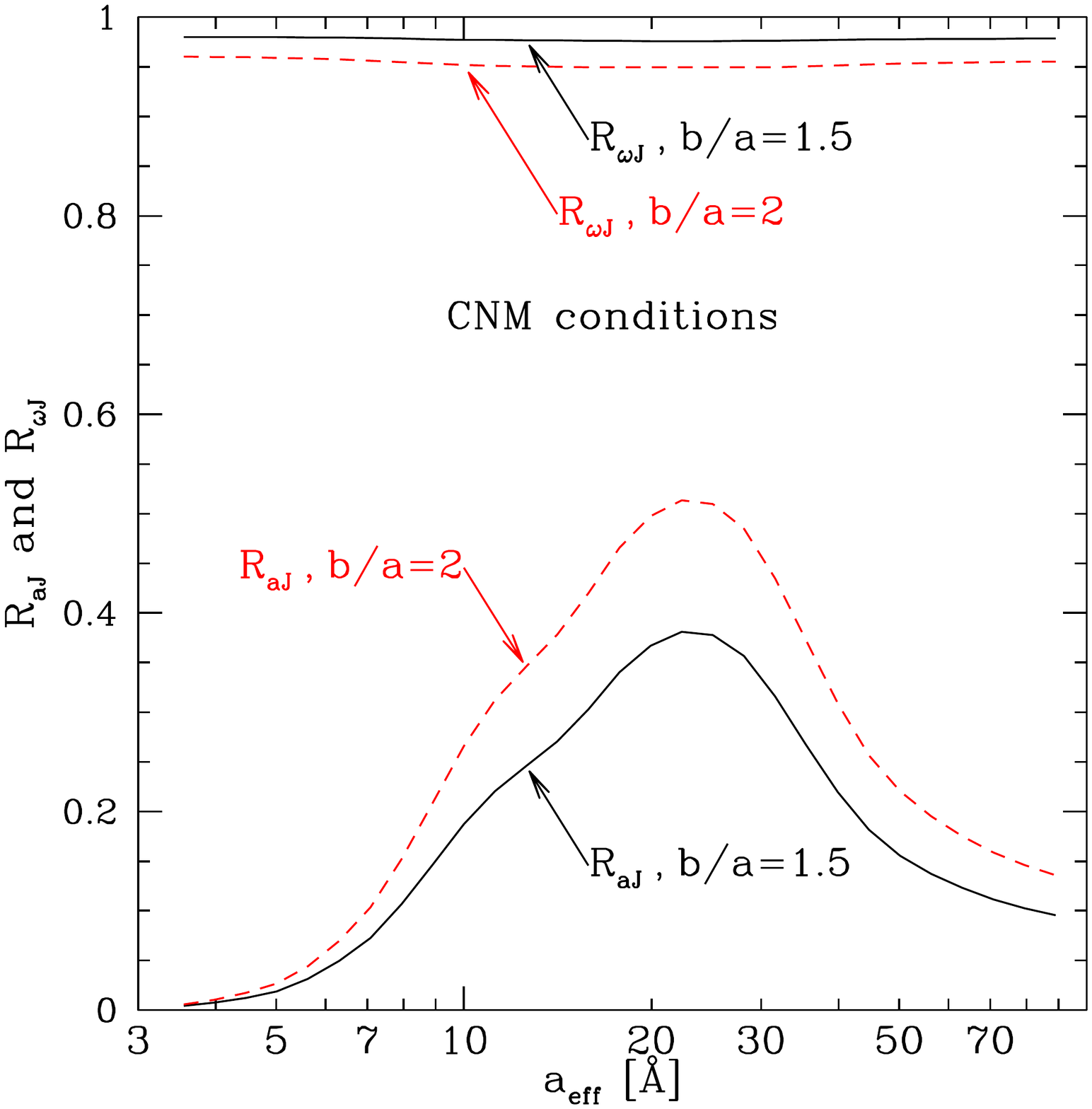}
\caption{\label{fig:R_aJ}\footnotesize
         Alignment factors $R_{\bahat_1\bJ}$ and $R_{\bomega\bJ}$
         for oblate spheroids with $b/a=1.5$ and 2.
         CNM conditions are assumed.
        }
\end{center}
\end{figure}

Recall that alignment of $\bahat_1$ with $\bJ$ occurs because the spinning
grain can reduce its kinetic energy by bringing
the axis of largest moment of inertia into alignment with $\bJ$.
Quantum suppression of dissipation will interfere with such alignment
in very small grains.
As an example, we consider oblate spheroids with axial ratio $b/a=2$, i.e.,
$B_v/A_v=2/[1+(a/b)^2]=1.6$.
The expectation value $\langle\cos^2\theta_{\bahat_1\bJ}\rangle$ is
evaluated using Eq.\ (\ref{eq:<cos2theta_aJ>}), using the rms
value of $J$ in Figure \ref{fig:frot}c.
Figure \ref{fig:R_aJ} shows $R_{\bahat_1\bJ}(a,J)$
as a function of grain radius $a$
for CNM excitation conditions (see Table \ref{tab:params}).
For radii $a\ltsim5\Angstrom$, the alignment of $\bahat_1$ with $\bJ$ is
minimal, although it becomes
appreciable for radii $10\Angstrom \ltsim a \ltsim 50\Angstrom$.

\subsection{Alignment of $\bomega$ with $\bJ$}

\begin{figure}[ht]
\begin{center}
\includegraphics[angle=0,width=5.0cm,
                 clip=true,trim=0.5cm 5.0cm 0.5cm 2.5cm]
{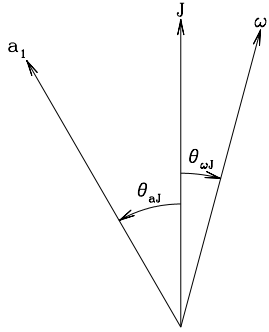}
\caption{\label{fig:geom}\footnotesize
The body axis $\bahat_1$, angular momentum $\bJ$, and
angular velocity $\bomega$.
$\bahat_1$ and $\bomega$ 
both nutate around the angular momentum vector $\bJ$.
$\bJ$ in turn precesses around the magnetic
field $\bB_0$ (not shown).}
\end{center}
\end{figure}

The electric dipole rotational emission from a single grain is 100\%
polarized if viewed from a direction perpendicular to the grain's
instantaneous angular velocity $\bomega$.
However, if $R_{\bahat_1\bJ}<1$, then $\bomega$ and $\bJ$ will not
be parallel, and both $\bahat_1$ and $\bomega$ will nutate
around $\bJ$ (see Figure \ref{fig:geom}).
For an oblate spheroid, 
the angle $\theta_{\bomega\bJ}$ between $\bomega$ and $\bJ$ is given by
\beq
\cos^2\theta_{\bomega\bJ} = 
\frac{[q + (1-q)\cos^2\theta_{\bahat_1\bJ}]^2} 
{q^2 + (1-q^2)\cos^2\theta_{\bahat_1\bJ}}
~~~,
\eeq
where $q\equiv I_\parallel/I_\perp \geq 1$.
In the limit of a sphere ($q\rightarrow 1$), $\bomega\parallel\bJ$,
and $\cos^2\theta_{\bomega\bJ}=1$ 
independent of the alignment of $\bahat_1$ with
$\bJ$.  For oblate particles with $q > 1$, misalignment of
$\bahat_1$ with $\bJ$ 
implies misalignment of $\bomega$ with $\bJ$, but this misalignment
is only slight.
For an ensemble of grains with angular momentum quantum number $J$, 
\beqa \label{eq:<cos2theta_omegaJ>}
\langle \cos^2\theta_{\bomega\bJ}\rangle
&=& \sum_{E>E_{\rm crit}} p_E
\sum_K 
\frac{\left[q+(1-q)K^2/J(J+1)\right]^2}
     {\left[q^2+(1-q^2)K^2/J(J+1)\right]}\,
p_K(J,T_E)
+ 
\\
&&\left(\sum_{E=0}^{E_{\rm crit}}p_E\right)
\sum_K 
\frac{\left[q+(1-q)K^2/J(J+1)\right]^2}
     {\left[q^2+(1-q^2)K^2/J(J+1)\right]}\,
p_K(J,T_{\rm crit})
~~~,
\eeqa
where $p_K(J,T)$ is given by Eq.\ (\ref{eq:p_K}),
and we define
\beq
R_{\bomega\bJ} \equiv 
\frac{3}{2}\left(\langle\cos^2\theta_{\bomega\bJ}\rangle -\frac{1}{3}\right)
~~~.
\eeq
Figure \ref{fig:R_aJ} shows $R_{\bomega\bJ}$ as a function of
grain size $a$, for $b/a=1.5$ ($q=1.385$) and $b/a=2$ ($q=1.6$).
We see that $R_{\bomega\bJ}>0.95$: 
$\bomega$ remains quite well-aligned with $\bJ$, even when 
$R_{\bahat_1\bJ}$ is small.

\section{\label{sec:rotpol}
         Polarization of Rotational Emission}

\begin{figure}[ht]
\begin{center}
\includegraphics[angle=0,width=8.0cm,
                 clip=true,trim=0.5cm 5.0cm 0.5cm 2.5cm]
{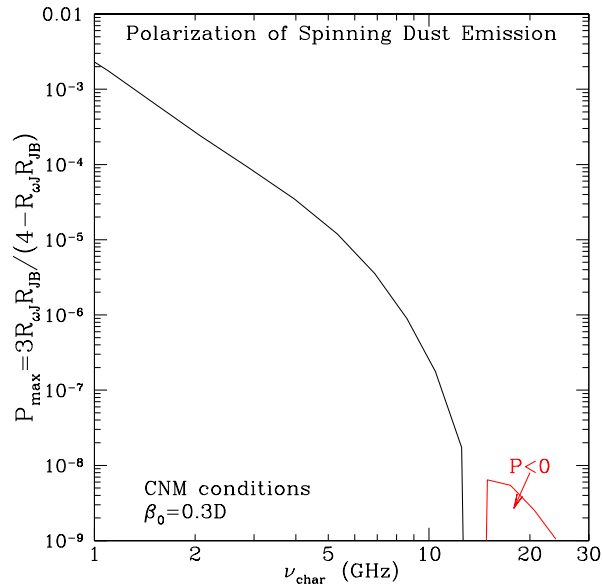}
\caption{\label{fig:fpol}\footnotesize
         Upper limit on polarization of rotational emission
         as a function of characteristic frequency $\nu_{\rm char}$.
         }
\end{center}
\end{figure}

If $R_{\bahat_1\bJ}< 1$, a spinning grain will
undergo nutation around $\bJ$.  A spinning grain will generally
have a significant magnetic moment antiparallel to $\bomega$ due
to the Barnett effect
\citep{Dolginov+Mytrophanov_1976,Purcell_1979}; 
this magnetic moment will cause $\bJ$
to precess around $\bB_0$ with a period that is short enough that
complete averaging over precession can be assumed.
If we assume that the grain has an electric dipole moment $\mu_\perp$
perpendicular to $\bomega$, then, after averaging over
rotation, nutation, and precession, one can show that the rotational emission
will have fractional polarization
\beq
P_{\rm rot}
= \frac{3R_{\bomega\bJ}R_{\bJ\bB}}{4-R_{\bomega\bJ}R_{\bJ\bB}}
\sin^2\Psi
~~~,
\eeq
where $\Psi$ is the angle between the viewing direction and
$\bB_0$.
Figure \ref{fig:fpol} shows the predicted polarization as a function
of frequency for viewing directions perpendicular to the
static magnetic field ($\sin^2\Psi=1$).
For each frequency $\nu$ we assume the emission to be dominated by
grains with $\nu_{\rm char}=\nu$.
At microwave frequencies $\nu>5\GHz$, 
the predicted polarizations are extremely small.
If the AME is rotational radiation from spinning dust grains, the
polarization should be negligible.

\section{\label{sec:pol opt-IR}
         Polarization of Extinction or Thermal Emission}

Nonspherical grains have absorption or scattering
cross sections that depend on the orientation of the grain relative
to the direction and
polarization of the incident radiation.  In the long wavelength
limit, the cross section depends on the direction of the polarization
$\bE$ relative to the grain body, but not on the direction of
propagation.  We assume this to be the case in the following
discussion.

\begin{figure}[ht]
\begin{center}
\includegraphics[angle=0,width=8.0cm,
                 clip=true,trim=0.5cm 5.0cm 0.5cm 2.5cm]
{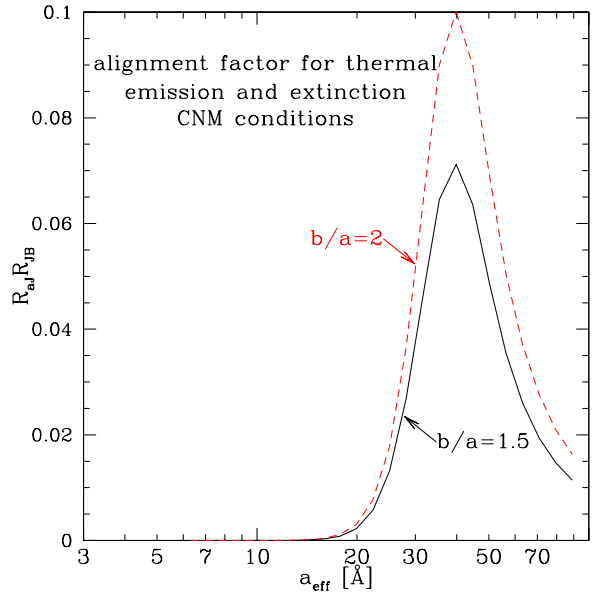}
\caption{\label{fig:R_aJ R_JB}
         Alignment factor $R_{\bahat_1\bJ}R_{\bJ\bB}$ for
         dichroic extinction or thermal emission.
         CNM conditions are assumed.
}
\end{center}
\end{figure}

We consider oblate spheroidal grains with symmetry axis $\bahat_1$.
Let $\behat$ be the direction of polarization, and let
$C_\parallel$ and $C_\perp$ be cross sections for $\behat\parallel\bahat_1$
and $\behat\perp\bahat_1$.  If we view the ensemble of grains from
a direction perpendicular to $\bB_0$, the ensemble of precessing and
nutating grains will have mean cross sections per particle
\beqa
\langle C\rangle_{\behat\parallel\bB}
&=& R_{\bahat_1\bJ}R_{\bJ\bB} C_\parallel + (1-R_{\bahat_1\bJ}R_{\bJ\bB})
\frac{(C_\parallel+2C_\perp)}{3}
\\
\langle C\rangle_{\behat\perp\bB}
&=& R_{\bahat_1\bJ}R_{\bJ\bB} C_\perp +  (1-R_{\bahat_1\bJ}R_{\bJ\bB})
\frac{(C_\parallel+2C_\perp)}{3}
~~~.
\eeqa
Thermal emission would then have polarization
\beq
P_{\rm th.em.} = \frac{\langle C\rangle_{\behat\perp\bB}-
          \langle C\rangle_{\behat\parallel\bB}}
         {\langle C\rangle_{\behat\perp\bB}+
         \langle C\rangle_{\behat\parallel\bB}}
=  \frac{3R_{\bahat_1\bJ}R_{\bJ\bB}(C_\perp-C_\parallel)}
{3R_{\bahat_1\bJ}R_{\bJ\bB} (C_\parallel+C_\perp) + 2(1-R_{\bahat_1\bJ}R_{\bJ\bB})(C_\parallel+2C_\perp)}
~~~.
\eeq
The polarization due to dichroic extinction by a column density $N$ of
dust grains is
\beqa
P_{\rm ext} &=& \tanh\left(N C_{\rm pol}\right)
\\
C_{\rm pol} &=& \frac{1}{2}\left( \langle C\rangle_{\behat\perp\bB}-
          \langle C\rangle_{\behat\parallel\bB} \right)
= \frac{1}{2}  R_{\bahat_1\bJ}R_{\bJ\bB} \left(C_\perp-C_\parallel\right)
~~~.
\eeqa
Thus for both emission and extinction the polarization is determined by
the product $R_{\bahat_1\bJ}R_{\bJ\bB}$ determining the degree of
grain alignment.

\section{\label{sec:discussion}
         Discussion}

\citet{Rouan+Leger+Omont+Giard_1992} considered IVRET in
spinning PAHs with $N\approx90$, and concluded that IVRET was sufficiently
rapid so that Eq.\ (\ref{eq:p_K}) should be a good approximation throughout
the cooldown following absorption of a starlight photon.
Quantum supression of dissipation in grains was reconsidered by
\citet[][hereafter LD00]{Lazarian+Draine_2000}, 
who noted that the energy $E_\ell$
of the lowest
vibrationally-excited state would be appreciable in small grains, and
argued that spin-lattice relaxation should be suppressed by
a factor $\propto\exp(-E_\ell/k\Tvib) \ll 1$, leading to suppression of
IVRET when $\Tvib$ drops below $\sim E_\ell/k$.
LD01 estimated that this would reduce the polarization to
$\sim2\%$ at 20\,GHz, and only $\sim$$0.5\%$ 
for particles small enough to spin at 30\,GHz.

\citet{Sironi+Draine_2009} revisited IVRET
in spinning PAHs.
They argued that when the separation $\Delta E$ of the vibrational
energy levels becomes larger than $\hbar\omega_{\rm rot}$, 
vibrational-rotational energy exchange will be suppressed.
For a PAH with $\sim$200 C atoms, they estimated that IVRET would
effectively cease when $\Tvib$ dropped below $\sim$$65\K$, leaving
the body axis only partially aligned with $\bJ$. 

In the present paper we have argued for a different criterion: 
that v-R energy transfer is suppressed when
$g_E\delta \Evib < 1$, where $\delta \Evib$ is the width of the energy states.
Equation (\ref{eq:psi}) is proposed as
an estimate for the quantum suppression factor $\psi_{\rm q}$.
This criterion leads to
$\psi_{\rm q}\ll1$ for very small grains
-- see Fig.\ \ref{fig:psi_dg}.
However, Equation (\ref{eq:psi}) probably 
{\it overestimates} the relaxation
rate when $g_E\delta E < 1$:
simply having a state $v_2$ available with the appropriate
energy does not ensure that the {\it coupling} from $v_1$ to $v_2$ 
will be fast, as
there may be other ``selection rules'' that must be satisfied to have
the energy transfer proceed at the ``classical'' rate.
Thus, the true suppression factor
$\psi_{\rm q}$ may be {\it smaller} than estimated from Eq.\ ({\ref{eq:psi}),
and the actual degree of polarization of rotational emission may
be even smaller than the already very small values estimated here, and
shown in Figure \ref{fig:rjb}.

The numerical values in Figures \ref{fig:frot} and \ref{fig:rjb}
were calculated for
nanoparticles with the properties of amorphous silicates.
However, other grain materials -- in particular, PAHs or metallic Fe --
would also have quantum suppression of alignment, 
qualitatively
similar to the silicate example shown here.
Thus, if the AME is dominated by rotational emission from nanoparticles
spinning at $\sim$$30\GHz$ frequencies, we expect the AME to
be negligibly polarized.

\begin{figure}[ht]
\begin{center}
\includegraphics[angle=0,width=8.0cm,
                 clip=true,trim=0.5cm 5.0cm 0.5cm 2.5cm]
                {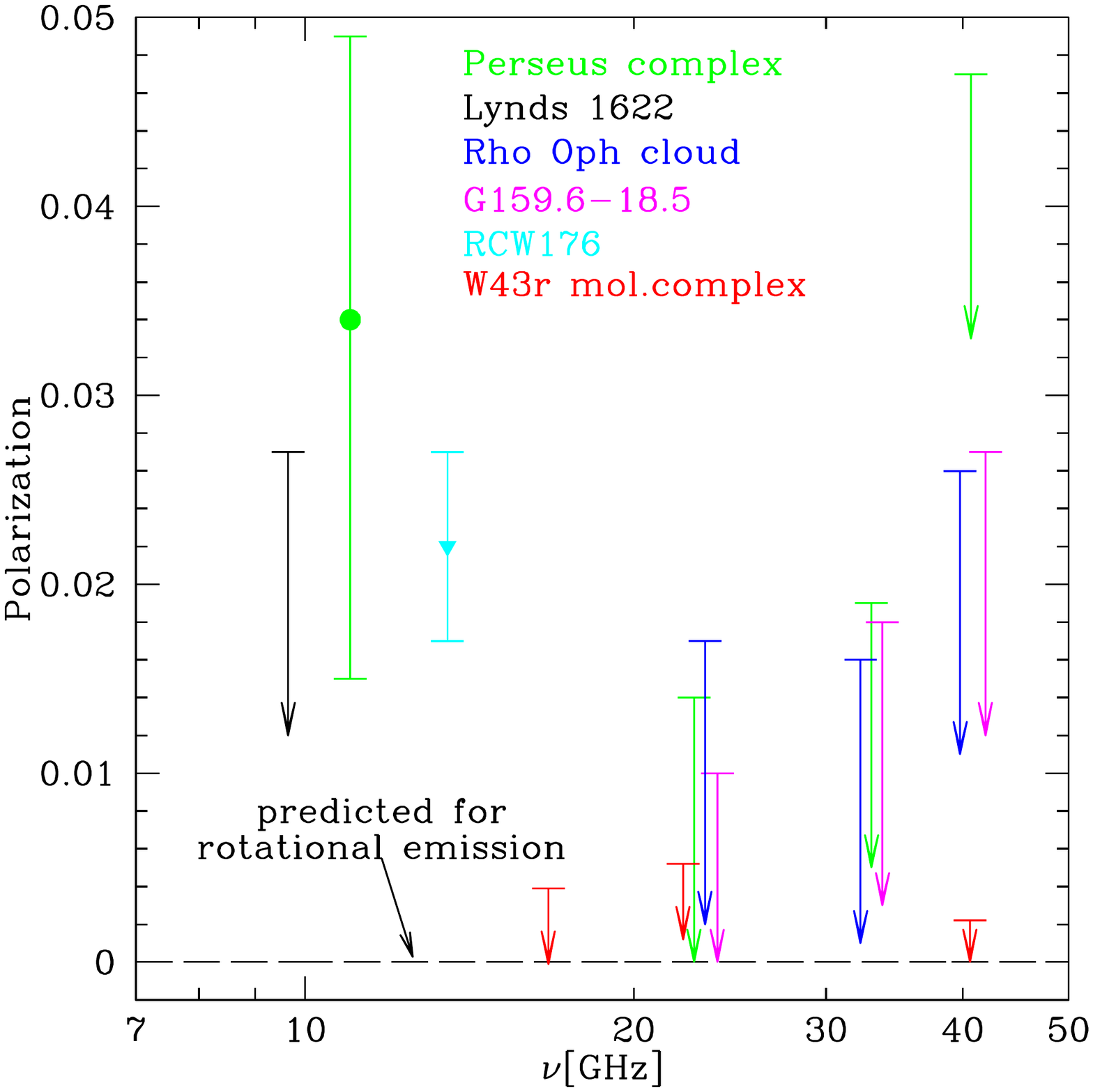}
\caption{\label{fig:pol_obs}\footnotesize
         Measured polarization of AME from
         the Perseus molecular complex 
         \citep{Battistelli+Rebolo+Rubino-Martin+etal_2006,
                Dickinson+Peel+Vidal_2011};
         the dark clouds Lynds 1622 \citep{Mason+Robishaw+Heiles+etal_2009}
         and $\rho\,$Oph \citep{Dickinson+Peel+Vidal_2011};
         the HII regions G159.6-18.5
         \citep{Lopez-Caraballo+Rubino-Martin+Rebolo+Genova-Santos_2011,
                Genova-Santos+Rubino-Martin+Rebolo+etal_2015}
         and RCW176
         \citep{Battistelli+Carretti+Cruciani+etal_2015};
         and the W43r molecular complex
         \citep{Genova-Santos+Rubino-Martin+Pelaez-Santos+etal_2016}.
         }
\end{center}
\end{figure}

Our prediction of negligible polarization for rotational emission
at microwave frequencies is consistent with published observations.
\citet{Battistelli+Rebolo+Rubino-Martin+etal_2006} measured the
total polarization to be
$P=3.4_{-1.9}^{+1.5}\%$ at 11\,GHz for the Perseus molecular complex, but
the observed polarization may be entirely due
to synchrotron emission.
At higher frequencies, the AME contribution from this region 
is consistent with zero,
as are observations of the AME from the Lynds 1622 dark cloud,
the $\rho\,$Oph cloud, and the HII region G159.6-18.5 --
see Figure \ref{fig:pol_obs}.
A polarization $P=2.2\pm0.5\%$ was reported for the HII region RCW176
and 13.5 GHz, but this could be also be due to synchrotron emission associated
with RCW176.
Most recently, \citet{Genova-Santos+Rubino-Martin+Pelaez-Santos+etal_2016}
have obtained very stringent upper limits of 0.39\%, 0.52\% and 0.22\%
for the AME polarization at 16.7, 22.7, and 40.6\,GHz.

At this time, our prediction of negligible polarization for rotational
emission at microwave frequencies is consistent with obervations.

Our modeling of the grain dynamics 
[Eq.\ (\ref{eq:dJpara2/dt},\ref{eq:dJperp2/dt})]
has implicitly assumed that the damping and excitation functions
$F$ and $G$ do not depend on the orientation of the grain angular momentum 
$\bJ$ --
the only dependence on orientation enters through the magnetic
dissipation term in Eq.\ (\ref{eq:dJperp2/dt}).
However, 
if the starlight illuminating the dust is anisotropic, and the grain
is appreciably nonspherical, the
photon absorption rate for a grain will depend on the orientation of
the grain relative to the starlight anisotropy.
If the grain axis $\bahat_1$ is aligned with $\bJ$, then the time-averaged
photon absorption rate will in principle depend on the orientation of
$\bJ$ and on the angle $\theta_{\bahat_1\bJ}$.
Recognizing that $\bahat_1$ will be nutating around $\bJ$, and $\bJ$ will
be precessing around $\bB_0$ (the grain will generally have a magnetic
moment), we do not expect $F$ and $G$ to have an appreciable dependence
on the orientation of $J$, but in principle it will not be zero unless
the starlight is isotropic.
Because the grain angular momentum
is changed by absorption of the starlight photon and the subsequent emission 
of infrared photons, the rotational distribution function
for the dust may develop a small degree of anisotropy as the result
of the starlight anisotropy.
Anisotropic starlight can
produce a small degree of polarization in the PAH emission features
from spinning PAHs in photodissociation regions 
\citep{Leger_1988,Sironi+Draine_2009}.  
For rotational emission this effect is expected to be slight.
If future observations find small but
nonzero polarization for the AME, the possible
contribution from this mechanism should be quantitatively evaluated.

It future observations find the AME to be polarized
with $P\gtsim0.01\%$ at $\nu\gtsim10\GHz$, it will
be evidence that the AME is not entirely rotational emission from
nanoparticles, or that starlight anisotropy has generated a small degree
of polarization in the rotational emission.

While quantum suppression of alignment in nanoparticles with
radii $a\ltsim 10\Angstrom$ will lead to effectively zero polarization
of any rotational emission at GHz frequencies, 
Davis-Greenstein
alignment 
is expected to be able to significantly align grains in the
$a\approx 30-50\Angstrom$ size range.
The minimal degree of polarization in the far-ultraviolet 
\citep{Martin+Clayton+Wolff_1999,Whittet_2004}
suggests that grains in this size range must be either nearly spherical or have
low abundance.

\section{\label{sec:summary}
         Summary}

The principal results of this paper are as follows:
\begin{enumerate}
\item Dissipation due to viscoelasticity or Barnett dissipation
in a spinning grain
is suppressed when the vibrational energy $\Evib$
falls below a critical value $E_\crit$: the grain's rotational
kinetic energy cannot be converted to vibrational energy because there are
no suitable vibrational states.
This suppresses alignment of the grain body axis $\bahat_1$ 
with the grain angular momentum $\bJ$.
\item 
Paramagnetic or ferromagnetic dissipation in a nanoparticle spinning in
a static magnetic field $\bB_0$ is also suppressed when 
$\Evib < E_\crit$.
This suppresses alignment of the grain angular momentum $\bJ$ with
the galactic magnetic field $\bB_0$, with the greatest suppression
for the smallest grains.
\item
For conditions typical of the neutral ISM, the rotational emission
from interstellar dust at frequencies $\nu > 1\GHz$ is expected to
be negligibly polarized, with $P< 10^{-6}$ for $\nu > 10\GHz$
(see Figure \ref{fig:fpol}).
If the anomalous microwave emission arises from spinning grains, it should
be essentially unpolarized, consistent with observations to date.
\item Ordinary paramagnetic dissipation should be able to align
dust grains in the $30-50\Angstrom$ size range.
The rapid fall-off in starlight polarization in the far-ultraviolet
suggests that grains in this size range are either nearly spherical
or contribute only a small fraction of the far-ultraviolet extinction.
\end{enumerate}
The above conclusions are not sensitive to the grain material, and apply
to rotational emission from spinning PAHs, nanosilicates, or nano-Fe
particles.
\acknowledgments
We thank
D.\ Gutkowicz-Krusin
for helpful discussions.
This work was supported in part by NSF grant AST-1408723,
and was carried out in part at the Jet Propulsion Laboratory,
California Institute of Technology, under a contract with the
National Aeronautics and Space Administration.

\bibliography{/u/draine/work/bib/btdrefs}

\end{document}